\documentclass[12pt]{article}
\usepackage{natbib}
\usepackage{graphicx}
\usepackage{setspace}
\usepackage{amssymb} 
\usepackage{amsmath} 
\usepackage{amsthm}
\usepackage{enumerate}
\usepackage{array}
\usepackage{multirow}
\usepackage{float}
\usepackage{caption}
\usepackage{subcaption}
\usepackage{natbib}
\usepackage{graphicx}

\PassOptionsToPackage{normalem}{ulem}
\usepackage{ulem}
\usepackage{xcolor}
\usepackage[
    colorlinks,
    linkcolor={blue!50!black},
    citecolor={blue!50!black},
    urlcolor={blue!80!black}]{hyperref}

\theoremstyle{plain}

\providecommand{\theoremname}{Theorem}

\usepackage[a4paper,left=0.7in,right=0.7in,top=1in,bottom=1in]{geometry}
\begin{document}
\sloppy

\title{Fusion of Tree-induced Regressions for Clinico-genomic Data}

 \author{\small{Jeroen M. Goedhart\footnote{Corresponding author, E-mail address: j.m.goedhart@amsterdamumc.nl} \textsuperscript{a}, Mark A. van de Wiel\textsuperscript{a}, Wessel N. van Wieringen\textsuperscript{a,b}, Thomas Klausch\textsuperscript{a}}}

\maketitle
\noindent\textsuperscript{a}\footnotesize{Department of Epidemiology and Data Science, Amsterdam Public Health Research Institute,
Amsterdam University Medical Centers Location AMC, Meibergdreef 9, Noord-Holland, the Netherlands}\\
\noindent\textsuperscript{b}\footnotesize{Department of Mathematics, Vrije Universiteit}, De Boelelaan 1081a, Noord-Holland, the Netherlands

\abstract{Cancer prognosis is often based on a set of omics covariates and a set of established clinical covariates such as age and tumor stage. Combining these two sets poses challenges. First, dimension difference: clinical covariates should be favored because they are low-dimensional and usually have stronger prognostic ability than high-dimensional omics covariates. Second, interactions:  genetic profiles and their prognostic effects may vary across patient subpopulations. Last, redundancy: a (set of) gene(s) may encode similar prognostic information as a clinical covariate. To address these challenges, we combine regression trees, employing clinical covariates only, with a fusion-like penalized regression framework in the leaf nodes for the omics covariates. The fusion penalty controls the variability in genetic profiles across subpopulations. We prove that the shrinkage limit of the proposed method equals a benchmark model: a ridge regression with penalized omics covariates and unpenalized clinical covariates. Furthermore, the proposed method allows researchers to evaluate, for different subpopulations, whether the overall omics effect enhances prognosis compared to only employing clinical covariates. In an application to colorectal cancer prognosis based on established clinical covariates and 20,000+ gene expressions, we illustrate the features of our method.}

\maketitle

\section{Introduction}
\label{sec:intro}
Because cancer is largely molecular in nature, biomedical studies
often employ omics derives for diagnosis and prognosis of the disease.
Along the measured omics covariates, well-established clinical covariates
such as age, smoking behavior, tumor stage or grade, and blood measures
are typically also available. These well-established covariates, sometimes
summarized by prognostic indices such as the International Prognostic
Index (IPI) and the Nottingham Prognostic Index (NPI), should be included
in the model of choice to render more accurate and stable predictions
\citep{SauerbreiClinOmics,bovelstat2009}. This manuscript presents a
method to deal with prognostic models based on omics derives and well-established
clinical risk factors. Such models are usually called clinico-genomic
models \citep{bovelstat2009}. 

As a motivating example, we consider a model that estimates
relapse-free survival of $914$ colorectal cancer (CRC) patients based on
a combination of expression levels of $21,292$ genes and clinical
covariates age, gender, tumor stage, and tumor site. Several considerations
should be taken into account for such a model. First, the large difference
in dimensionality: the omics data are high-dimensional, so shrinkage
is required for these covariates, whereas only few clinical covariates
are available. Second, it is expected that on average a clinical covariate
adds more to prognosis than an omics covariate. Third, interactions
between the clinical and omics covariates may be present. For example,
stage \mbox{I} and stage \mbox{IV} patients may strongly differ in
their genetic profile and its effect on the outcome, which ideally
should be taken into account. In addition, for some clinically-based
subpopulations, e.g. Stage \mbox{IV} patients that are older than
$80,$ the overall omics effect may hardly improve prognosis. A model
that finds such patterns provides valuable information on the added
benefit of measuring relatively costly omics covariates.

To address the aforementioned challenges, we present FusedTree, a
novel clinico-genomic model. The main idea is to fit a regression
tree using solely the clinical covariates and, subsequently, fitting
linear models in the leaf nodes of the tree using the omics covariates.
The regression tree automatically finds potential interaction terms
between clinical covariates and it naturally handles ordinal (e.g.
tumor stage) and categorical data. Furthermore, subsamples in the
different nodes belong to well-defined clinically-based subpopulations,
which therefore allows for easy assessment of the benefit of omics
data for prognosis of a particular subpopulation. Because trees are
less-suited for continuous variables (e.g. age), we also include such
variables additively with unpenalized linear effects in the model.

Each node has its own omics-based regression and hence interactions
between clinical covariates and omics covariates are modeled. To control
the interaction strength, we incorporate a fusion-like penalty into
the omics-based regression estimators. Specifically, this penalty
shrinks the omics effect estimates in the different nodes to each
other. Furthermore, coupling the regressions in the different nodes
stabilizes effect size estimation. We also include a standard penalty
to each omics-based regression to accommodate the high-dimensionality
of omics data. The intercepts of the linear models in the nodes, which
correspond to the effects of the clinical covariates, are left unpenalized
to account for their established predictive power. This overall shrinkage
procedure renders a unique ridge-based penalized likelihood framework
which can be optimized efficiently for (very) large numbers of omics
covariates. Furthermore, we prove that the strength of the proposed
fusion-like penalty interpolates between a fully interactive model,
in which the omics-based regression in each node is estimated freely,
and a standard ridge regression model, in which no clinical-omics
interactions are present. We opt for ridge penalties instead of lasso
penalties because ridge often outperforms lasso in prediction, as
we will also show in simulations, and because omics applications are
rarely sparse \citep{BOYLE2017}.

The remainder of this work is organized as follows. We start by reviewing
related models and alternative strategies to clinico-genomic modeling
in Section \ref{sec:Related-work} and \ref{subsec:Alternative-strategies-for},
respectively. Section \ref{sec:Method} deals with a detailed description
of the methodology of FusedTree, which handles continuous, binary,
and survival response. Subsequently, we illustrate the benefits of
FusedTree compared to other models in simulations (Section \ref{sec:Simulations}).
We then apply FusedTree to the aforementioned colorectal cancer prognosis
study in Section \ref{sec:Application}. We conclude with a summary
and a discussion in Section \ref{sec:Discussion}.

\subsection{Related models}
\label{sec:Related-work}
FusedTree is a type of model-based partitioning, first suggested by \citet{ZeileisModelBasedPartitioning}. Model-based partitioning
recursively tests for parameter instability of model covariates, in
our case the omics covariates, with respect to partitioning covariates,
in our case the clinical covariates. A splitting rule is created with
the partitioning covariate showing the largest model parameter instability.
This is done recursively until all model parameter instability is
resolved within some tolerance level. FusedTree has important distinctions
compared to the model-based partitioning. First, we do not optimize
the tree and the linear models in the leafs jointly, but instead first
fit a tree with just the clinical covariates and then conditional
on the tree the linear models in the leafs. Optimizing the tree structure
for only the clinical covariates acknowledges their established predictive
power. Second, as mentioned above, we regularize the fit to account
for high-dimensionality and we link the regressions in the different
nodes to obtain more stable estimates.

Model-based partitioning is a varying coefficients model
\citep{VaryingCoefModels}. Such a model allows the effects of a set
of predictors to vary with a different set of predictors/effect modifiers. A relevant example is glinternet \citep{Glinternet}, a model that allows for sparsely incorporating interactions between a low-dimensional covariate set and a (potentially) high-dimensional covariate set.  \citet{VaryingCofClinOm} proposed
modeling interactions between omics covariates and a linear combination of the clinical
covariates by smoothing splines. Omics effects and omics-clinical
covariate interactions are estimated using lasso-based
penalties. This model, however, does not allow for nonlinear clinical covariate effects, and is, combined with lasso penalties, arguably
better suited for variable selection than for prediction. 

\subsection{Alternative strategies for clinico-genomic data}
\label{subsec:Alternative-strategies-for}
Other models addressing some of the challenges of clinico-genomic
data may be divided in two groups: linear models and nonlinear models.
For linear models, a simple solution is to employ a regularization
framework in which the the clinical covariates are penalized differently
(or not penalized at all) compared to the omics covariates. Examples
implementing this idea are IPF-Lasso \citep{LassoIPF} employing lasso
penalization \citep{LassoTibshirani}, and multistep elastic net \citep{Established_vs_Unestablished}
employing elastic net penalization \citep{ElasticNet}. Another linear
approach is boosting ridge regression \citep{BoostingClinOmics},
in which, at each boosting step, a single covariate is updated according
to a penalized likelihood criterion with a large penalty for the omics
covariates and no penalty for the clinical covariates. 
Downsides of linear clinico-genomic models compared to FusedTree are
1) the clinical part may possess nonlinearities which may be estimated
fairly easily because the clinical part is usually low-dimensional,
2) clinical-omics covariate interactions are less straightforwardly
incorporated, especially when part of the clinical data is ordinal/categorical.

For nonlinear models, tree-based methods such as random forest \citep{Breiman2001},
gradient boosting \citep{GBFriedman}, and Bayesian additive regression
trees (BART) \citep{chipmanBART2010} are widely used. To incorporate
a clinical-omics covariate hierarchy into tree-based methods, the
prior probabilities of covariates being selected in the splitting
rules may be adjusted, e.g. by upweighting the clinical covariates.
Block forest considers a random forest with covariate-type-specific
selection probabilities, which are estimated by cross-validation \citep{BlockForest2019}.
EB-coBART considers the same strategy as Block Forests, but employs
BART as base-learner and estimates the covariate-type-specific selection
probabilities using empirical Bayes \citep{goedhart2023codata}. A
downside of sum-of-trees models is their complexity, which is arguably
too large to reliably estimate effects of high-dimensional omics covariates.
Additionally, interpreting such models is more challenging compared
to FusedTree (and penalized regression models). We illustrate how FusedTree may be used for interpretation in Section \ref{sec:Application}.

\section{FusedTree}
\label{sec:Method}

\subsection{Set-up}
\label{sec:Set-up}
Let data $\left\{ y_{i},\boldsymbol{x}_{i},\boldsymbol{z}_{i}\right\} _{i=1}^{N}$
consist of $N$ observations, indexed by $i,$ of a response $y_{i},$
an omics covariate vector $\boldsymbol{x}_{i}\in\mathbb{R}^{p}$ having
elements $x_{ij},$ and clinical covariate vector $\boldsymbol{z}_{i}\in\mathbb{R}^{q}$
having elements $z_{il}.$ We collect the clinical and omics covariate
measurements in design matrices $\boldsymbol{Z}=\left(\boldsymbol{z}_{1}^{\top},\ldots,\boldsymbol{z}_{N}^{\top}\right)^{\top}\in\mathbb{R}^{N\times q},$
and $\boldsymbol{X}=\left(\boldsymbol{x}_{1}^{\top},\ldots,\boldsymbol{x}_{N}^{\top}\right)^{\top}\in\mathbb{R}^{N\times p},$
respectively. We assume that $\boldsymbol{z}_{i}$ is low-dimensional
and that $\boldsymbol{x}_{i}$ is high-dimensional, i.e. $q<N<p.$
We further assume normalized $\boldsymbol{x}_{i}$ (zero mean and
standard deviation equal to $1$). We present our method for continuous
$y_{i}$ and briefly describe differences with binary and survival
response for which full details are found in supplementary Sections
1 and 2, respectively. 

\begin{figure}
\includegraphics[bb=13bp 150bp 955bp 530bp,clip,width=\textwidth]{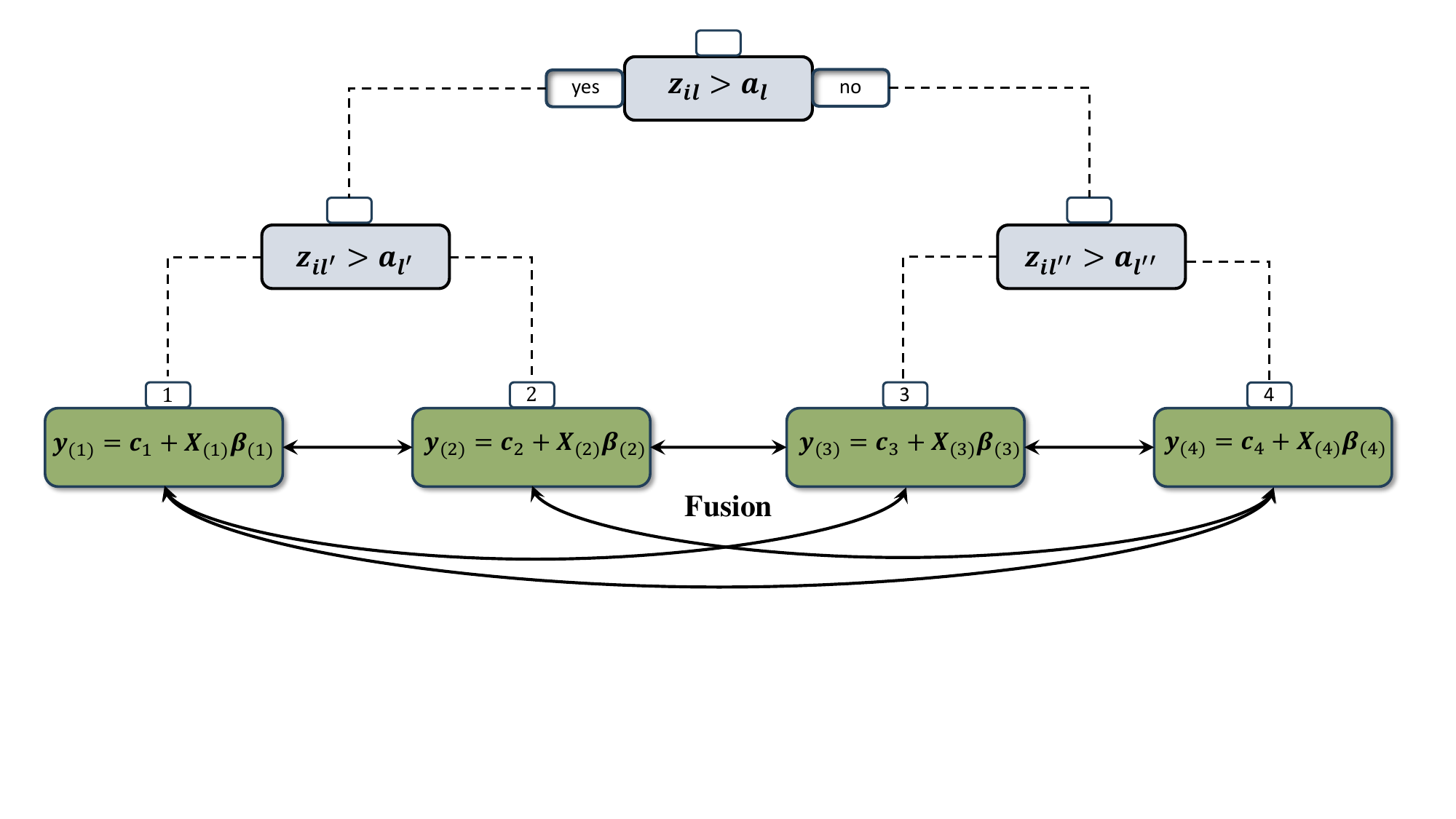}\caption{\label{fig:Set-up-of-Method}Set-up of FusedTree. In each
leaf node $m$ ($m=1,\ldots,4$ in this example), we fit a linear
regression using $n_{m}$ samples with omics covariates $\boldsymbol{X}_{(m)}$
and an intercept $\boldsymbol{c}_{m}$. The intercept contains the
(potentially nonlinear) clinical information. The regression in leaf
node $m$ borrows information from the other leaf nodes by linking
the regressions (indicated with $\protect\longleftrightarrow)$ through
fusion penalty \eqref{eq:fusionPenalty}.}
\end{figure}

In prediction, we consider $y_{i}=f\left(\boldsymbol{x}_{i},\boldsymbol{z}_{i}\right)+\epsilon_{i},$ with error $\epsilon_{i}$ an iid unobserved random variable with $\mathbb{E}[\epsilon_{i}]=0,$
and we aim to estimate a function $f\left(\cdot\right)$
that accurately predicts $y_{i}.$ Clinical covariates $\boldsymbol{z}_{i}$
should often be prioritized above $\boldsymbol{x}_{i}$ in $f\left(\cdot\right)$
because of their established predictive value compared to omics covariates.
To acknowledge the difference in predictive power and dimensions of
the two types of covariates, we propose to combine regression trees
with linear regression models in the leaf nodes. The regression trees
are estimated using the clinical covariates $\boldsymbol{z}_{i}$
only, thereby accounting for possible nonlinearities and interactions.
Subsequently, the linear regressions in the leaf nodes are fitted
using the omics covariates $\boldsymbol{x}_{i}$ (including an intercept
term to account for $\boldsymbol{z}_{i}$). Thus, we fit cluster-specific
linear regressions using omics covariates with the clusters defined
in data-driven fashion by fitting a tree with the clinical covariates.
Our method, which we call FusedTree, is summarized in Figure \ref{fig:Set-up-of-Method}.

\subsection{Regression Trees}
\label{sec:RegressionTrees}
We fit regression trees using the CART algorithm \citep{CART1984}
implemented in the R package \href{https://cran.r-project.org/web/packages/rpart/index.html}{\texttt{rpart}}.
CART clusters the clinical covariates $\boldsymbol{z}$ by $M$ nonoverlapping
(hyper)rectangular regions $\boldsymbol{R}=\left\{ R_{m}\right\} _{m=1}^{M}$
in the clinical covariate space $\mathcal{Z}.$ Clusters $R_{m}$
correspond to the leaf nodes of the tree. CART then predicts $y_{i}$
by assigning constants $c_{m},$ combined in vector $\boldsymbol{c}=\left(c_{1},\ldots,c_{M}\right)^{T}\in\mathbb{R}^{M},$
to the corresponding $R_{m}.$ Thus, we have the following prediction
model$:$
\begin{equation}
f\left(\boldsymbol{c},\boldsymbol{R};\boldsymbol{z}_{i}\right)=\sum_{m=1}^{M}c_{m}I\left(\boldsymbol{z}_{i}\in R_{m}\right),\label{eq:RegTree}
\end{equation}
with $I\left(\cdot\right)$ the indicator function. 

Regions/leaf nodes $R_{m}$ are defined by a set of binary splitting
rules $\left\{ z_{il}>a_{l}\right\} ,$ with each rule representing
an internal node of the tree. The rules are found in greedy fashion
by computing the split that renders the largest reduction in average
node impurity, which we quantify by the mean square error for continuous
$y_{i}$ and the Gini index for binary $y_{i}.$ For survival response,
we use the deviance of the full likelihood of a proportional hazards
model \citep{SurvTreesCrowley} as is implemented in the R package
\href{https://cran.r-project.org/web/packages/rpart/index.html}{\texttt{rpart}}.

To prevent overfitting, we post-prune the tree by penalizing the number
of terminal nodes $M$ with pruning hyperparameter $\kappa$. The
best $\kappa$ is determined using $K$-fold cross-validation \citep{CART1984}.
We also consider a minimal sample size in the nodes of $30$ to avoid
too few samples for the omics-based regressions.

\subsection{Model}
\label{sec:Model}
FusedTree adds omics-based regressions
to the leaf-node-specific constants $c_{m}:$
\begin{equation}
y_{i}\mid\boldsymbol{R}=f\left(\boldsymbol{c},\boldsymbol{\beta};\boldsymbol{x}_{i},\boldsymbol{z}_{i}\right)+\epsilon_{i}=\sum_{m=1}^{M}\left(c_{m}+\boldsymbol{x}_{i}^{\top}\boldsymbol{\beta}_{(m)}\right)I\left(\boldsymbol{z}_{i}\in R_{m}\right)+\epsilon_{i},\label{eq:Model}
\end{equation}
with $\boldsymbol{\beta}_{(m)}\in\mathbb{R}^{p}$
the leaf-node-specific omics regression parameter vectors having elements
$\beta_{j(m)}.$ All omics parameter vectors are combined in the vector
$\boldsymbol{\beta}=\left(\boldsymbol{\beta}_{(1)}^{\top},\ldots,\boldsymbol{\beta}_{(M)}^{\top}\right)^{\top}\in\mathbb{R}^{Mp}$.
Model \eqref{eq:Model} treats the fitted tree structure defined by
$\boldsymbol{R}$ as fixed. Specifically, we first determine $\boldsymbol{R}$
using $\boldsymbol{z}_{i}$ only and then consider \eqref{eq:Model}.
Parameters $c_{m}$ and $\boldsymbol{\beta}_{(m)}$
will be estimated jointly. Model \eqref{eq:Model} defines $y_{i}$
as a combination of a clinically-based intercept $c_{m}$, which is
usually nonlinear in $\boldsymbol{z}_{i}$, and a linear omics part
$\boldsymbol{x}_{i}\boldsymbol{\beta}_{(m)}.$
Because $\boldsymbol{\beta}_{(m)}$ is leaf-node-specific,
model \eqref{eq:Model} also incorporates interactions between $\boldsymbol{x}_{i}$
and $\boldsymbol{z}_{i}.$ 

For binary response, $y_{i}\in\left\{ 0,1\right\},$ we consider
$y_{i}\mid\boldsymbol{R},\boldsymbol{x}_{i},\boldsymbol{z}_{i}\sim\textrm{Bern}\left\{ \exp\left(f(\cdot)\right)/\left[\exp\left(f(\ldotp)\right)+1\right]\right\} ,$
while for survival response, we consider a Cox proportional hazards
model \citep{CoxPropHaz}: $h\left(t\mid\boldsymbol{R},\boldsymbol{x}_{i},\boldsymbol{z}_{i}\right)=h_{0}\left(t\right)\exp\left(f(\cdot)\right),$
with $f(\cdot)$ defined as in model \eqref{eq:Model}, and $h_{0}\left(t\right)$
the baseline hazard function.

To recast model \eqref{eq:Model} in matrix notation, we define leaf-node
specific data\\ $\left(\boldsymbol{1}_{n_{m}},\boldsymbol{X}_{(m)},\boldsymbol{y}_{(m)}\right)=\left\{ 1,\boldsymbol{x}_{i},y_{i}\right\} _{i:\boldsymbol{z}_{i}\in R_{m}},$
with $\boldsymbol{1}_{n_{m}}\in\mathbb{R}^{n_{m}}$ a vector of all
ones indicating the leaf-node-specific intercept for node $m$ (clinical
effect), and omics $\boldsymbol{X}_{(m)}\in\mathbb{R}^{n_{m}\times p}$
and response $y^{(m)}\in\mathbb{R}^{n_{m}}$ observations in leaf
node $m$. 

Next, we collect the data of all $M$ leaf nodes in the block-diagonal
omics matrix $\tilde{\boldsymbol{X}}\in\mathbb{R}^{N\times Mp},$
the block-diagonal leaf-node-intercept-indicator matrix $\tilde{\boldsymbol{U}}\in\mathbb{R}^{N\times M},$
and response vector $\tilde{\boldsymbol{y}}\in\mathbb{R}^{N}:$
\[
\tilde{\boldsymbol{U}}=\begin{pmatrix}\boldsymbol{1}_{n_{1}} & \boldsymbol{0}_{n_{1}} & \cdots & \boldsymbol{0}_{n_{1}}\\
\boldsymbol{0}_{n_{2}} & \boldsymbol{1}_{n_{2}} & \boldsymbol{\ddots} & \vdots\\
\vdots & \ddots & \ddots & \boldsymbol{0}_{n_{M-1}}\\
\boldsymbol{0}_{n_{M}} & \boldsymbol{\cdots} & \boldsymbol{0}_{n_{M}} & \boldsymbol{1}_{n_{M}}
\end{pmatrix},\quad\tilde{\boldsymbol{X}}=\begin{pmatrix}\boldsymbol{X}_{(1)} & \boldsymbol{0}_{n_{1}\times p} & \cdots & \boldsymbol{0}_{n_{1}\times p}\\
\boldsymbol{0}_{n_{2}\times p} & \boldsymbol{X}_{(2)} & \boldsymbol{\ddots} & \vdots\\
\vdots & \ddots & \ddots & \boldsymbol{0}_{n_{M-1}\times p}\\
\boldsymbol{0}_{n_{M}\times p} & \boldsymbol{\cdots} & 0 & \boldsymbol{X}_{(M)}
\end{pmatrix},\quad\tilde{\boldsymbol{y}}=\begin{pmatrix}\boldsymbol{y}_{(1)}\\
\boldsymbol{y}_{(2)}\\
\vdots\\
\boldsymbol{y}_{(M)}
\end{pmatrix},
\]
with $\boldsymbol{0}$ a vector/matrix with all zeros. We then rewrite
model \eqref{eq:Model} to 
\begin{equation}
\tilde{\boldsymbol{y}}=\underbrace{\tilde{\boldsymbol{U}}\boldsymbol{c}}_{\textrm{clinical}}\quad+\underbrace{\tilde{\boldsymbol{X}}\boldsymbol{\beta}}_{\textrm{omics}\times\textrm{clinical}}+\quad\boldsymbol{\epsilon},\label{eq:Matrix=000020model}
\end{equation}
where we absorb the dependence/conditioning on $\boldsymbol{R}$ of
Model \eqref{eq:Matrix=000020model} in the $\tilde{\cdot}$ notation.
Recall the clinical effect vector $\boldsymbol{c}=\left(c_{1},\ldots,c_{M}\right)^{T},$
which collects the leaf-node specific intercepts. 

\subsection{Penalized estimation}
\label{sec:PenEstimation}
We jointly estimate clinical effects $\boldsymbol{c}$ by $\hat{\boldsymbol{c}}$
and omics effects \textbf{$\boldsymbol{\beta}$} by $\hat{\boldsymbol{\beta}}$
using penalized least squares optimization. We leave $\hat{\boldsymbol{c}}$
unpenalized to account for the established predictive power of the
clinical covariates $\boldsymbol{z}_{i}$. We penalize $\hat{\boldsymbol{\beta}}$
by 1) the standard ridge penalty \citep{RidgeHoerl} controlled by
hyperparameter $\lambda>0$ to accommodate high-dimensional settings
and 2) a fusion-type penalty controlled by hyperparameter $\alpha>0$
to shrink the interactions between the covariates $\boldsymbol{x}_{i}$
and $\boldsymbol{z}_{i}.$ This fusion-type penalty shrinks elements
$\beta_{(1)j},\beta_{(2)j},\ldots,\beta_{(M)j},$
which represent the effect sizes of omics covariate $j$ in the different
leaf nodes/clinical clusters, to their shared mean. More fusion shrinkage
implies more similar $\beta_{(1)j},\beta_{(2)j},\ldots,\beta_{(M)j},$
which reduces the interaction effects between omics and clinical covariates.
Furthermore, the fusion-type penalty ensures that each leaf node regression
is linked to the other leaf node regressions, which allows for information
exchange.

Specifically, estimators $\hat{\boldsymbol{c}}$ and $\hat{\boldsymbol{\beta}}$
are found by
\begin{equation}
\hat{\boldsymbol{c}},\,\hat{\boldsymbol{\beta}}=\underset{\boldsymbol{c},\boldsymbol{\beta}}{\textrm{arg max}}\:L\left(\boldsymbol{c},\boldsymbol{\beta};\tilde{\boldsymbol{U}},\tilde{\boldsymbol{X}},\tilde{\boldsymbol{y}}\right)-\lambda\boldsymbol{\beta}^{T}\boldsymbol{\beta}-\alpha\boldsymbol{\beta}^{\top}\boldsymbol{\Omega}\boldsymbol{\beta},\label{eq:PenalizedLik}
\end{equation}
with $L\left(\boldsymbol{c},\boldsymbol{\beta};\tilde{\boldsymbol{U}},\tilde{\boldsymbol{X}},\tilde{\boldsymbol{y}}\right)=\left\Vert \tilde{\boldsymbol{y}}-\tilde{\boldsymbol{U}}\boldsymbol{c}-\tilde{\boldsymbol{X}}\boldsymbol{\beta}\right\Vert _{2}^{2}$
the least squares estimator, $\lambda\boldsymbol{\beta}^{\top}\boldsymbol{\beta}$
the standard ridge penalty, and fusion-type penalty 
\begin{equation}
\alpha\boldsymbol{\beta}^{\top}\boldsymbol{\Omega}\boldsymbol{\beta}=\alpha\sum_{m=1}^{M}\sum_{j=1}^{p}\left(\beta_{(m)j}-\bar{\beta}_{j}\right)^{2},\quad\bar{\beta}_{j}=\frac{1}{M}\sum_{m=1}^{M}\beta_{(m)j},\label{eq:fusionPenalty}
\end{equation}
with fusion matrix $\boldsymbol{\Omega}\in\mathbb{R}^{Mp\times Mp}.$
Penalty \eqref{eq:fusionPenalty} shrinks the effects of omics covariate
$j$ in the different nodes to their shared mean $\bar{\beta}_{j},$
which reduces the interaction effect sizes between clinical and omics
covariates. Importantly, this shared mean is not specified in advance,
but is also learned from the data. This shrinkage approach is related
to ridge to homogeneity proposed by
\citet{RidgeHomogeneity}. Penalty \eqref{eq:fusionPenalty}, however,
only shrinks specific elements of $\boldsymbol{\beta}$ to a shared
value, whereas ridge to homogeneity
shrinks all elements to a shared value. 

Matrix $\boldsymbol{\Omega}$ has a block diagonal structure with
identical blocks after reshuffling the elements of \textbf{$\boldsymbol{\beta}$
}(and corresponding columns of $\tilde{\boldsymbol{X}}).$ By redefining
$\boldsymbol{\beta}=\left(\beta_{(1)1},\beta_{(2)1}\dots,\beta_{(M)1},\beta_{(1)2},\dots\beta_{(M)2},\dots,\beta_{(1)p},\dots\beta_{(M)p}\right)^{\top}$,
the fusion matrix equals $\boldsymbol{\Omega}=\boldsymbol{I}_{p\times p}\bigotimes\left(\boldsymbol{I}_{M\times M}-\frac{1}{M}\boldsymbol{1}_{M\times M}\right),$
with $\boldsymbol{1}_{M\times M}$ a matrix with all elements equal
to $1.$ Matrix $\boldsymbol{\Omega}$ is nonnegative definite and
therefore, after including $\lambda\boldsymbol{\beta}^{\top}\boldsymbol{\beta},$
the optimization in \eqref{eq:PenalizedLik} has a unique solution.

Solving optimization \eqref{eq:PenalizedLik} renders, as derived
by \citet{10.1093/jrsssc/qlad041}, the following estimators
for $\boldsymbol{c}$ and $\boldsymbol{\beta}:$
\begin{align}
\hat{\boldsymbol{c}} & =\left\{ \tilde{\boldsymbol{U}}^{\top}\left[\tilde{\boldsymbol{X}}\left(\lambda\boldsymbol{I}_{Mp\times Mp}+\alpha\boldsymbol{\Omega}\right)^{-1}\tilde{\boldsymbol{X}}^{\top}+\boldsymbol{I}_{N\times N}\right]^{-1}\tilde{\boldsymbol{U}}\right\} ^{-1}\nonumber \\
 & \times\tilde{\boldsymbol{U}}^{\top}\left[\tilde{\boldsymbol{X}}\left(\lambda\boldsymbol{I}_{Mp\times Mp}+\alpha\boldsymbol{\Omega}\right)^{-1}\tilde{\boldsymbol{X}}^{\top}+\boldsymbol{I}_{N\times N}\right]^{-1}\tilde{\boldsymbol{y}}\nonumber \\
\hat{\boldsymbol{\beta}} & =\left(\tilde{\boldsymbol{X}}^{\top}\tilde{\boldsymbol{X}}+\lambda\boldsymbol{I}_{Mp\times Mp}+\alpha\boldsymbol{\Omega}\right)^{-1}\tilde{\boldsymbol{X}}^{\top}\left(\tilde{\boldsymbol{y}}-\tilde{\boldsymbol{U}}\hat{\boldsymbol{c}}\right).\label{eq:Estimators}
\end{align}
By defining $\boldsymbol{W}=\left[\tilde{\boldsymbol{X}}\left(\lambda\boldsymbol{I}_{Mp\times Mp}+\alpha\boldsymbol{\Omega}\right)^{-1}\tilde{\boldsymbol{X}}^{\top}+\boldsymbol{I}_{N\times N}\right]^{-1},$
estimator $\hat{\boldsymbol{c}}=\left(\tilde{\boldsymbol{U}}^{\top}\boldsymbol{W}\tilde{\boldsymbol{U}}\right)^{-1}\tilde{\boldsymbol{U}}^{\top}\boldsymbol{W}\tilde{\boldsymbol{y}}$
is recognized as the weighted least squares estimator with weights
related to the variation in $\tilde{\boldsymbol{X}}.$
This reformulation implies that observations with a large variation
in omics covariates are downweighted in their contribution to clinical
effects estimator $\hat{\boldsymbol{c}}$.

The shrinkage limits of \eqref{eq:Estimators}, as we derive in Supplementary Section
4, equal
\begin{align}
\lim_{\lambda\rightarrow\infty}\hat{\boldsymbol{c}} & =\left(\tilde{\boldsymbol{U}}^{\top}\tilde{\boldsymbol{U}}\right)^{-1}\tilde{\boldsymbol{U}}^{\top}\tilde{\boldsymbol{y}},\quad\lim_{\lambda\rightarrow\infty}\hat{\boldsymbol{\beta}}=\boldsymbol{0}_{Mp},\nonumber \\
\lim_{\alpha\rightarrow\infty}\hat{\boldsymbol{c}} & =\left\{ \tilde{\boldsymbol{U}}^{\top}\left[\boldsymbol{X}\left(\frac{1}{\lambda M}\boldsymbol{I}_{p\times p}\right)\boldsymbol{X}^{\top}+\boldsymbol{I}_{N\times N}\right]^{-1}\tilde{\boldsymbol{U}}\right\} ^{-1}\tilde{\boldsymbol{U}}^{\top}\left[\boldsymbol{X}\left(\frac{1}{\lambda M}\boldsymbol{I}_{p\times p}\right)\boldsymbol{X}^{\top}+\boldsymbol{I}_{N\times N}\right]^{-1}\tilde{\boldsymbol{y}}\nonumber \\
\underset{\alpha\rightarrow\infty}{\lim}\hat{\boldsymbol{\beta}} & =\left[\left(\boldsymbol{X}^{\top}\boldsymbol{X}+\lambda M\boldsymbol{I}_{p\times p}\right)^{-1}\boldsymbol{X}^{\top}\left(\tilde{\boldsymbol{y}}-\tilde{\boldsymbol{U}}\hat{\boldsymbol{c}}\right)\right]\ast\boldsymbol{1}_{M\times N}.\label{eq:ShrinkLimits}
\end{align}
Thus, $\underset{\lambda\rightarrow\infty}{\lim}$ reduces $\hat{\boldsymbol{c}}$
to the standard normal equation, and shrinks the omics effect sizes
to zero, as expected. Limit $\underset{\alpha\rightarrow\infty}{\lim}$
reduces the FusedTree estimators in \eqref{eq:Estimators} to a standard
ridge regression with the original omics matrix $\boldsymbol{X}\in\mathbb{R}^{N\times p},$
and penalty $\lambda M\boldsymbol{I}_{p\times p}.$ Note that the
penalty is a factor $M$ (number of leaf nodes) larger to account
for having $Mp$ parameter estimates instead of $p$. The notation
$\ast$ indicates the column-wise Kronecker product \citep{KhatriRhao}
with $\boldsymbol{1}_{M\times N},$ which ensures that each entry
$j$ of the standard ridge estimator is repeated $M$ times. We show
regularization paths, i.e. estimators \eqref{eq:Estimators} as a
function of fusion penalty $\alpha$ for several fixed values of $\lambda$
in Supplementary Section 5 (Figure S2)
for a simulated data example. 

For binary $y_{i}\in\left\{ 0,1\right\} ,$ we consider optimizing a penalized Bernoulli likelihood with identical  penalization terms $\lambda\boldsymbol{\beta}^{\top}\boldsymbol{\beta}$
and $\alpha\boldsymbol{\beta}^{\top}\boldsymbol{\Omega}\boldsymbol{\beta}.$
The penalized likelihood is optimized using iterative
re-weighted least squares (IRLS). For survival response, we use a
penalized proportional hazards model in which the regression parameters
are found by optimizing the full penalized likelihood using IRLS, similarly
to binary $y_{i}\in\left\{ 0,1\right\}$ \citep{HouwelingenCox}.
Full details are found in Supplementary Sections 1
and 2.

\subsection{Efficient hyperparameter tuning}
\label{sec:HypTuning}
We tune hyperparameters $\lambda$ and $\alpha$ by optimizing a $K$-fold
cross-validated predictive performance criterion. We partition the
data into $K$ non-overlapping test folds $\varGamma_{k},$ with $\varGamma_{k}$
a set of indices $\left\{ i\right\} _{i\in\varGamma_{k}}$ indicating
which observations from data $\mathcal{D}$ belong to $\varGamma_{k}.$
The number of samples in each $\varGamma_{k}$ should be as equal
as possible. Furthermore, for FusedTree, the folds are stratified
with respect to the tree-induced clinical clusters. For binary response,
we also balance the folds.

For test fold $\varGamma_{k},$ we then estimate the model parameters
on the training fold $\left(-\varGamma_{k}\right)$ and estimate the
performance on $\varGamma_{k}.$ We then aim to find $\lambda=\hat{\lambda},\,\alpha=\hat{\alpha}$
such that the average performance over the $K$ folds is optimized.
For continuous response, we use the mean square error as performance
measure, and hence we solve: 
\begin{equation}
\hat{\lambda},\,\hat{\alpha}=\underset{\lambda,\,\alpha}{\textrm{arg min}}\frac{1}{K}\sum_{k=1}^{K}\left\Vert \tilde{\boldsymbol{y}}_{\varGamma_{k}}-\tilde{\boldsymbol{U}}_{\varGamma_{k}}\hat{\boldsymbol{c}}_{-\varGamma_{k}}\left(\lambda,\alpha\right)-\tilde{\boldsymbol{X}}_{\varGamma_{k}}\hat{\boldsymbol{\beta}}_{-\varGamma_{k}}\left(\lambda,\alpha\right)\right\Vert _{2}^{2},\quad\textrm{subject to}\quad\lambda,\,\alpha>0.\label{eq:hyperparameterTuning}
\end{equation}
Optimization \eqref{eq:hyperparameterTuning} is computationally intensive
because a $Mp\times Mp$ matrix has to inverted, costing $\mathcal{O}\left((Mp)^{3}\right),$
repeatedly according to \eqref{eq:Estimators} until \eqref{eq:hyperparameterTuning}
is at a minimum.

To solve \eqref{eq:hyperparameterTuning} in computationally more
efficient fashion, we may evaluate the linear predictors $\tilde{\boldsymbol{U}}_{\varGamma_{k}}\hat{\boldsymbol{c}}_{-\varGamma_{k}}$
and $\tilde{\boldsymbol{X}}_{\varGamma_{k}}\hat{\boldsymbol{\beta}}_{-\varGamma_{k}}$
without having to directly evaluate $\hat{\boldsymbol{c}}_{-\varGamma_{k}}$
and $\hat{\boldsymbol{\beta}}_{-\varGamma_{k}},$
as was shown by \citet{Multiridge}. For our penalized regression setting
with penalties $\lambda\boldsymbol{\beta}^{\top}\boldsymbol{\beta}$
and $\alpha\boldsymbol{\beta}^{\top}\boldsymbol{\Omega}\boldsymbol{\beta}$,
\citet{10.1093/jrsssc/qlad041} showed, for general nonnegative $\boldsymbol{\Omega}$,
how to efficiently compute $\tilde{\boldsymbol{U}}_{\varGamma_{k}}\hat{\boldsymbol{c}}_{-\varGamma_{k}}$
and $\tilde{\boldsymbol{X}}_{\varGamma_{k}}\hat{\boldsymbol{\beta}}_{-\varGamma_{k}},$
which only requires repeated operations with relatively small matrices
of dimension $N-\left|\varGamma_{k}\right|.$ 

Prior to these repeated operations, we compute the eigendecomposition
$\boldsymbol{\Omega}=\boldsymbol{V}_{\varOmega}\boldsymbol{D}_{\varOmega}\boldsymbol{V}_{\varOmega}^{T},$
with eigenbasis $\boldsymbol{V}_{\varOmega}$ and diagonal eigenvalue
matrix $\boldsymbol{D}_{\varOmega},$ and the matrix $\tilde{\boldsymbol{X}}'=\tilde{\boldsymbol{X}}\boldsymbol{V}_{\varOmega}\left(\lambda\boldsymbol{I}_{p\times p}+\alpha\boldsymbol{D}_{\varOmega}\right)^{-\frac{1}{2}}$
once. For $\boldsymbol{\Omega}=\boldsymbol{I}_{p\times p}\bigotimes\left(\boldsymbol{I}_{M\times M}-\frac{1}{M}\boldsymbol{1}_{M\times M}\right),$
the eigenbasis equals $\boldsymbol{V}_{\varOmega}=\boldsymbol{I}_{p\times p}\bigotimes\boldsymbol{V}_{A},$
with $\boldsymbol{V}_{A}$ the eigenbasis for $\boldsymbol{A}=\left(\boldsymbol{I}_{M\times M}-\frac{1}{M}\boldsymbol{1}_{M\times M}\right),$
and the eigenvalues are $\boldsymbol{D}_{\varOmega}=\boldsymbol{I}_{p\times p}\bigotimes\boldsymbol{D}_{A},$
with $\boldsymbol{D}_{A}$ the eigenvalues of $\boldsymbol{A}.$ Computing
$\boldsymbol{V}_{A}$ and $\boldsymbol{D}_{A}$ only costs $\mathcal{O}\left(M^{3}\right),$
while computing $\tilde{\boldsymbol{X}}'$ requires $\mathcal{O}\left((Mp)^{2}\right).$ 

To summarize, tuning $\lambda$ and $\alpha$ requires a single operation
quadratic in $Mp,$ after which only operations in dimension $N$
are required. For the typical $Mp\gg N,$ this means a significant
reduction in computational time compared to a naive evaluation of
\eqref{eq:hyperparameterTuning}. 

Full details on how to compute $\tilde{\boldsymbol{U}}_{\varGamma_{k}}\hat{\boldsymbol{c}}_{-\varGamma_{k}}$
and $\tilde{\boldsymbol{X}}_{\varGamma_{k}}\hat{\boldsymbol{\beta}}_{-\varGamma_{k}}$
are found in Supplementary Section 3
(including binary and survival response).

\subsection{Inclusion of linear clinical covariate effects}
\label{subsec:Inclusion-of-linear}
A single regression tree may model interaction/nonlinear effects,
but is less suited for modeling additive effects and continuous covariates.
Ensemble methods such as random forest \citep{Breiman2001} and gradient
boosted trees \citep{GBFriedman} (partly) solve this issue by combining
multiple trees additively. However, combining FusedTree with ensemble
methods will greatly increase computational time and more importantly,
the model will be harder to interpret. We therefore propose to additively
incorporate the clinical covariates $\boldsymbol{z}_{i}$ linearly
in the model as well. These linear effects will be absorbed in the
clinical design matrix $\tilde{\boldsymbol{U}}.$ We only incorporate
continuous covariates, categorical/ordinal covariates are only used
for tree fitting. The inclusion of linear clinical effects hardly
increases the number of covariates considering the dimension of the
omics design matrix $\tilde{\boldsymbol{X}}.$

\subsection{Test for the added value of omics effects in the leaf nodes}
\label{subsec:NodeRemoval}
In some instances, (a combination of) clinical covariates may (partly)
encode the same predictive information as (a combination of) omics
covariates. For FusedTree, this implies that in node $m$, the clinical
intercept $c_{m}$ contains most predictive power and estimating the
omics effects $\boldsymbol{\beta}_{(m)}$ is
not necessary. Omitting omics effects in some of the nodes renders
a simpler model. Furthermore, the nodes that only require a clinical
effect do not impact tuning of the fusion parameter $\alpha,$
which may therefore lead to improved tuning of $\alpha$ and the subsequent
estimation of $\boldsymbol{\beta}_{(m)}$ in
the nonempty nodes. Last and most importantly, because the nodes correspond
to well-defined and easy to understand clinically-based clusters,
FusedTree provides valuable information on the benefit of measuring
relatively costly omics covariates for diagnosis or prognosis of patient
subpopulations.

In principle, we may evaluate all $2^{M}$ possibilities of including/excluding
$\boldsymbol{\beta}_{(m)}$ in FusedTree and
then select the simplest model that predicts well. However, this quickly
becomes computationally intensive for large $M$. To balance between
model simplicity, predictive performance, and computational feasibility,
similarly as in backward selection procedures, we suggest the following
heuristic strategy, summarized by bullet points:
\begin{itemize}
\item In each node separately, we test whether the omics covariates add
to the explained variation of the response. For the hypothesis test,
we employ the global test implemented in the R package
\href{https://www.bioconductor.org/packages/release/bioc/html/globaltest.html}{\texttt{globaltest}}
\citep{GlobalTest}. Shortly, the test computes a score statistic that
quantifies how much the sum of all omics covariates combined add to
the explained variation of the response compared to solely using an
intercept. In Supplementary Section 6, we provide more detail
on the global test method in the context of FusedTree. The global
test renders a p-value for each node $m$: $p_{1},\ldots,p_{M}$,
which guide a greedy search for the best model.
\item We order the p-vales from largest (suggesting small added explained
variation of omics covariates) to smallest. We denote the ordered
p-value vector by $\boldsymbol{p}^{ord}.$
\item We fit several FusedTree models, guided by $\boldsymbol{p}^{ord}.$
We start by fitting the full FusedTree model, i.e. without any omics
effects removed. Then, we remove $\boldsymbol{\beta}_{(m')}$ and
$\boldsymbol{X}_{(m')}$ associated with the first element of $\boldsymbol{p}^{ord}$
and re-estimate model \eqref{eq:Model}. Next, we remove $\boldsymbol{\beta}_{(m')},\boldsymbol{\beta}_{(m'')}$
and $\boldsymbol{X}_{(m')},\boldsymbol{X}_{(m'')},$ associated with
the first two elements of $\boldsymbol{p}^{ord}$ and re-estimate
model \eqref{eq:Model}. We do so until all omics effects are removed
rendering a total of $M+1$ models. 
\item The model that balances between predictive power, estimated on an
independent test set, and simplicity, i.e. for how many nodes omics
covariates are present, should be preferred. Selecting the final FusedTree
model may be context dependent. For example, when omics measurements
are costly, stronger preference for simpler models is advisable. As
a rule of thumb, we suggest opting for the simplest model that is
performs maximally $2\%$ less than the model with the best test performance.
Because we only evaluate $M+1$ models, with typically $M<5,$ the
optimism bias introduced by this method is minimal. 
\end{itemize}

\section{Simulations}
\label{sec:Simulations}

We conduct three simulation experiments with different functional
relationships $f=\left(f_{1}f_{2},f_{3}\right)$ between continuous
response $y=f\left(\boldsymbol{z},\boldsymbol{x}\right)+\epsilon_{i},$
with $\epsilon_{i}\sim\mathcal{N}\left(0,1\right),$ and clinical
covariates $\boldsymbol{z}\in\mathbb{R}^{5}$ and omics covariates
$\boldsymbol{x}\in\mathbb{R}^{500}$ to showcase FusedTree:
\begin{enumerate}
\item Interaction. We specify
$f_{1}$ inspired by model \eqref{eq:Model}: 
\begin{align*}
f_{1}\left(\boldsymbol{x},\boldsymbol{z},\boldsymbol{\beta}\right) & =I\left(z_{1}\leq2.5\right)I\left(z_{2}\leq\frac{1}{2}\right)\left(-10+8\boldsymbol{x}_{1:125}^{\top}\boldsymbol{\beta}_{1:125}\right)\\
 & +I\left(z_{1}\leq2.5\right)I\left(z_{2}>\frac{1}{2}\right)\left(-5+2\boldsymbol{x}_{1:125}^{\top}\boldsymbol{\beta}_{1:125}\right)\\
 & +I\left(z_{1}>2.5\right)I\left(z_{3}\leq\frac{1}{2}\right)\left(5+\frac{1}{2}\boldsymbol{x}_{1:125}^{\top}\boldsymbol{\beta}_{1:125}\right)\\
 & +I\left(z_{1}>2.5\right)I\left(z_{3}>\frac{1}{2}\right)\left(10+\frac{1}{8}\boldsymbol{x}_{1:125}^{\top}\boldsymbol{\beta}_{1:125}\right)\\
 & +\boldsymbol{x}_{126:500}^{\top}\boldsymbol{\beta}_{126:500.}+5z_{4}.
\end{align*}
Clinical covariates are simulated according to 
Thus, $f_{1}$ is a tree with $4$ leaf nodes, defined by clinical covariates, with different
linear omics models in the leaf nodes for $25\%$ of the omics covariates.
The remaining $75\%$ of the omics covariates has a constant effect
size. 
\item Full Fusion. We specify $f_{2}$
by two separate parts, a nonlinear clinical part and a linear omics
part:
\[
f_{2}\left(\boldsymbol{x},\boldsymbol{z},\boldsymbol{\beta}\right)=15\sin\left(\pi z_{1}z_{2}\right)+10\left(z_{3}-\frac{1}{2}\right)^{2}+2\exp\left(z_{4}\right)+2z_{5}+\boldsymbol{x}^{\top}\boldsymbol{\beta}.
\]
 Clinical and omics covariates do not interact, so FusedTree should
benefit from a large fusion penalty.
\item Linear. In this experiment, we specify
$f_{3}$ by a separate linear clinical and a linear omics part:
\[
f_{3}\left(\boldsymbol{x},\boldsymbol{z},\boldsymbol{c},\boldsymbol{\beta}\right)=\boldsymbol{z}^{\top}\boldsymbol{c}+\boldsymbol{x}^{\top}\boldsymbol{\beta}.
\]
 Again, FusedTree should benefit from a large fusion penalty.
\end{enumerate}

\begin{figure}[!ht]
\includegraphics[width=\textwidth]{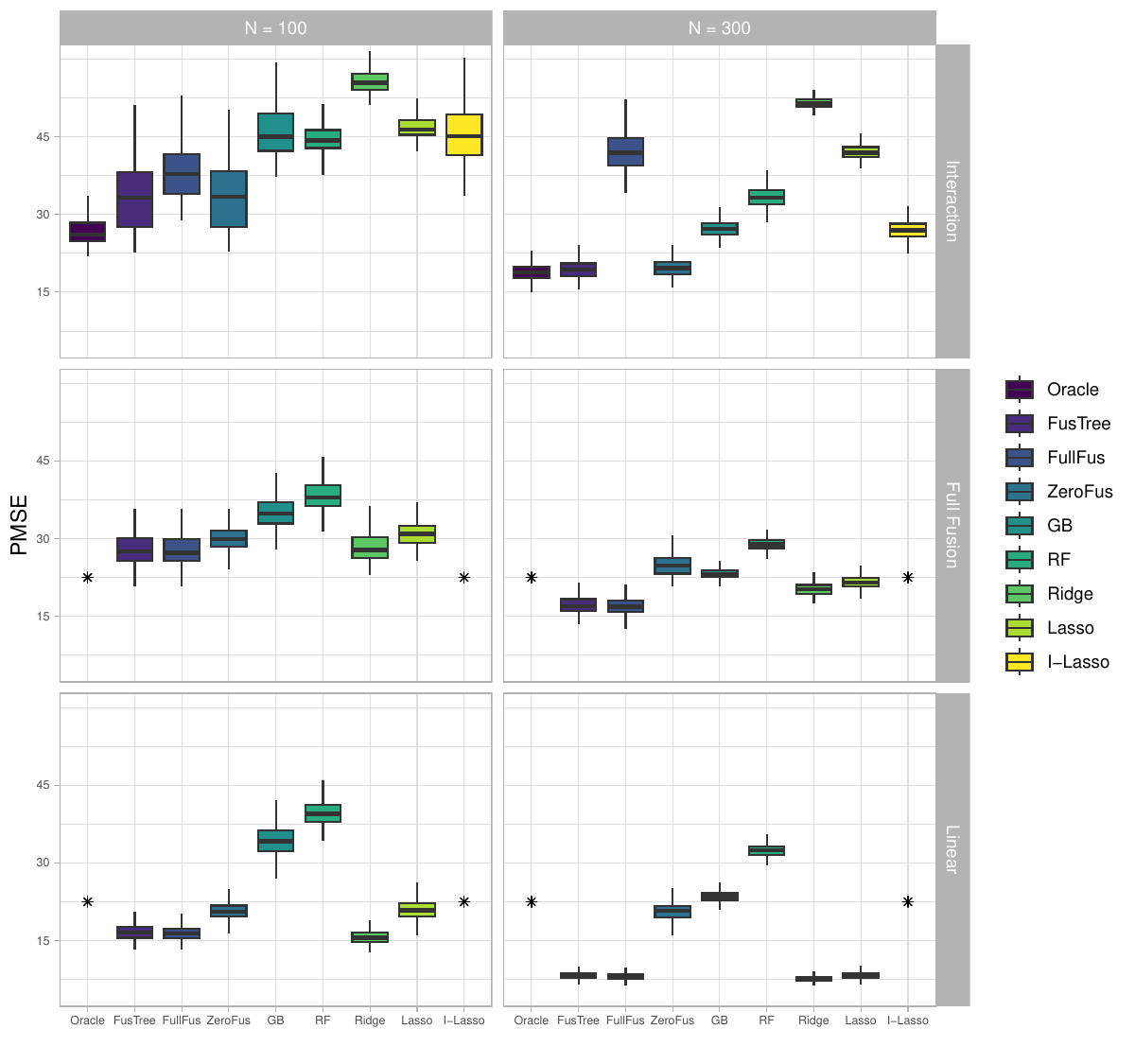}\caption{\label{fig:SimFigure}Boxplots of the prediction mean square
errors of several prediction models across $500$ simulated data sets
for the Interaction(top), Full Fusion (middle), and Linear (bottom) simulation experiment. For all experiments, we consider $N=100$
(left) and $N=300$ (right). The oracle prediction model is only considered
for the Interaction experiment ($\ast$
indicates that oracle model boxplots are missing for the Full
Fusion and Linear experiment).
We do not depict results for ridge regression in the Interaction
experiment because its PMSE's fall far outside the range of the PMSE's
of the other models (indicated by $\boldsymbol{\uparrow}$). Outliers
of boxplots are not shown.}
\end{figure}

Full descriptions of the experiments are found in Supplementary Section  7.
Shortly, for each experiment, we consider two simulation settings:
$N=100$ and $N=300.$ For each experiment and for each setting, we
simulate $N_{\textrm{sim}}=500$ data sets with $i=1,\ldots,N,$ and
clinical covariates $\boldsymbol{z}_{il}\sim\textrm{Unif}\left(0,1\right),$
for $l=1,\ldots,5$, and omics covariates $\boldsymbol{x}_{i}\sim\mathcal{N}\left(\boldsymbol{0}_{p},\boldsymbol{\Sigma}_{p\times p}\right),$
with $p=500,$ and correlation matrix $\boldsymbol{\Sigma}_{p\times p}$
set to the estimate of a real omics data set \citep{best2015rna}.
We simulate elements $j$ of the omics effect regression parameter
vector by $\beta_{1},\ldots,\beta_{p}\sim\textrm{Laplace}(0,\theta),$
with scale parameter $\theta.$ The Laplace distribution is the prior
density for Bayesian lasso regression and ensures many effect sizes
that are close to zero. 

To each data set, we fit FusedTree (FusTree) and several competitors:
ridge regression and lasso regression with unpenalized $\boldsymbol{z}_{i}$
and penalized $\boldsymbol{x}_{i},$ random forest (RF) , and gradient
boosted trees (GB). To assess the benefit of tuning fusion penalty
$\alpha,$ we also fit FusedTree with $\alpha=0$ (ZeroFus), and Fully
FusedTree (FulFus). Fully FusedTree jointly estimates a separate clinical
part, defined by the estimated tree, and a separate linear omics part that
does not vary with respect to the clinical covariates, which corresponds to FusedTree with $\alpha=\infty$ as shown by \eqref{eq:ShrinkLimits}. For the Interaction experiment, we also include an oracle tree model. This model knows
the tree structure in advance and only estimates the regression parameters
in the leaf nodes and tunes $\lambda$ and $\alpha$. For all FusedTree-based
models, we include all continuous clinical covariates $\boldsymbol{z}_{i}$
linearly in the regression model, as explained in Section \ref{subsec:Inclusion-of-linear}.
We quantify the predictive performance by the prediction mean square
error (PMSE), i.e. $N_{\textrm{test}}^{-1}\sum_{i=1}^{N_{\textrm{test}}}\left(y_{i}-\hat{y}_{i}\right)^{2},$
estimated on an independent test set with $N_{\textrm{test}}=5,000.$

FusedTree has a lower prediction mean square error (PMSE) compared
to the linear models ridge and lasso regression for the Interaction
and Full Fusion experiment because nonlinear
clinical effects are better captured by FusedTree (Figure \ref{fig:SimFigure}).
For the Linear experiment, FusedTree performs only marginally worse
than ridge regression, and has a slightly smaller PMSE compared to
lasso, even though omics effect sizes $\boldsymbol{\beta}$ were drawn
from a lasso prior. These findings suggest that 1) ridge penalties
are better suited for prediction compared to lasso penalties and 2)
the inclusion of linear clinical effects (Section \ref{subsec:Inclusion-of-linear})
to the tree ensures that linear clinical-covariate-response relationships
are only marginally better approximated by ridge regression compared
to FusedTree. FusedTree clearly outperforms nonlinear models random
forest and gradient boosted trees for all experiments. Gradient boosting
has a lower PMSE than random forest because we simulated mainly low-order
interactions, which can be better approximated by shallow trees, as
is the case for gradient boosting. 

The experiments also show a clear benefit of having a fusion-type
penalty whose strength is tuned by $\alpha.$ For the Full Fusion
and Linear experiment, for which no interactions between clinical
and omics covariates are present, FusedTree, which tunes $\alpha,$
performs nearly identical to an \textit{a priori} fully fused model,
which corresponds to setting $\alpha\rightarrow\infty$ in advance.
Furthermore, FusedTree performs better than FusedTree without the
fusion-type penalty, i.e. when we set $\alpha=0$ in advance. This
finding suggests the benefit of borrowing information across leaf
nodes. For the Interaction experiment, FusedTree benefits from tuning
$\alpha,$ such that interactions between clinical and omics covariates
may be modeled, by showing a clearly better performance compared to
the fully fused model. 

\section{Application}
\label{sec:Application}

\subsection{Description of the data}
We apply FusedTree to a combination of $4$ publicly available cohorts
consisting of $914$ colorectal adenocarcinoma patients with microsatellite
stability (MSS) for which we aim to predict relapse-free survival
based on $21,292$ gene expression covariates and clinical covariates:
age, gender, tumor stage (4-leveled factor), and the site of the
tumor (left versus right). In addition, a molecular clustering covariate
called consensus molecular subtype \citep{CMS_ColCancer} is available.
This clustering covariate, having four levels related to gene pathways,
mutation rates, and metabolics, is an established prognostic factor
and hence we include it to the clinical covariate set. The combined
cohorts are available as a single data set in the R package \href{https://bioconductor.org/packages/release/data/experiment/html/mcsurvdata.html}{\texttt{mcsurvdata}}. 

Patients with missing response values were omitted, rendering a final
data set with $N=845$ and $253$ events. Missing values in the clinical
covariate set were imputed using a single imputation with the R package
\href{https://cran.r-project.org/web/packages/mice/index.html}{\texttt{mice}} \citep{MCIE}.

\subsection{Model fitting and evaluation}
We fit FusedTree and several competitors to the data. We consider
FusedTree with and without post removal of omics effects in the nodes
as described in Section \ref{subsec:NodeRemoval}. We incorporate continuous covariate age linearly in FusedTree, as explained in Section \ref{subsec:Inclusion-of-linear}.
We fit the tree with a minimal leaf node sample size of $30$ and
we prune the tree and tune penalty parameters $\lambda$ and $\alpha$
using 5-fold CV. 

As competitors, we consider tree-based methods random survival forest
\cite{Ishwaran2008} implemented in the R package \href{https://cran.r-project.org/web/packages/randomForestSRC/index.html}{\texttt{randomforestSRC}},
gradient boosted survival trees implemented in the R package \href{https://cran.r-project.org/web/packages/gbm/index.html}{\texttt{gbm}},
and block forest \citep{BlockForest2019}, a random survival forest
which estimates separate weights for the clinical and omics covariates. 

For the linear competitor models, we consider a cox proportional hazards
model with only the clinical covariates, and we consider
lasso and ridge cox regression, both implemented in the R package
\href{https://cran.r-project.org/web/packages/glmnet/index.html}{\texttt{glmnet}} \citep{GlMNetCox},
with unpenalized clinical covariates and penalized omics covariates.
To favor clinical covariates more strongly, we also consider fitting
a cox proportional hazards model with only clinical covariates, and,
subsequently, fitting the residuals of this model using penalized
regression with only the omics covariates, as proposed by
\citet{OffSet_Strategy}. This residual approach, however, performs worse than jointly estimating the clinical (unpenalized) and the omics (penalized) effects, and we therefore do not show its results.
We do not consider CoxBoost \citep{BoostingClinOmics}, mentioned in Section \ref{subsec:Alternative-strategies-for},
because publicly available software was missing.

To evaluate the fit of all different models, we estimate the test
performance. To do so, we split the data set in a training set ($N_{\textrm{train}}=676$)
on which we fit the models, and a test set ($N_{\textrm{test}}=169$)
on which we estimate the performance. We show survival curves of the
training and test response in supplementary Figure S9. As performance
metrics, we consider the robust (against censoring distribution) concordance
index (C-index) \citep{RobustCStatistics} and the time-dependent area
under the curve (t-AUC) \citep{tAUC} using a cut-off of five years. 

We investigate the effect of the number of omics covariates $p$
on the fitted models. Therefore, we consider $p_{\textrm{sel}}=\left\{ 500,5000,21292\,\textrm{(all)}\right\} $
and select the $p_{\textrm{sel}}$ genes with the largest variance.

\subsection{Results and downstream analysis}
\begin{figure}
    \centering
    \includegraphics[width=\textwidth]{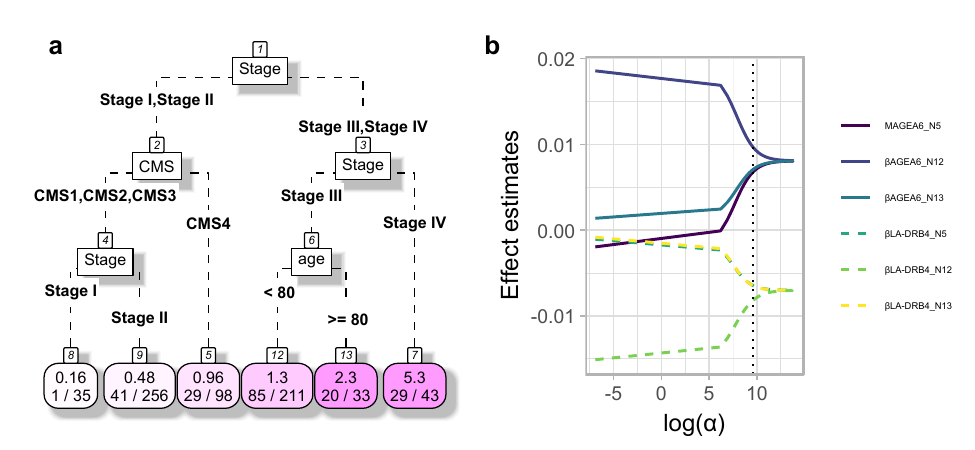}
    \caption{(\textbf{a}) The estimated survival tree of FusedTree. In the leaf
nodes, the relative death rate (top) and the
number of events/node sample size (bottom) are depicted. The plot is produced using the
R package \href{https://cran.r-project.org/web/packages/rpart.plot/index.html}{\texttt{rpart.plot}}.
(\textbf{b}) Regularization paths as a function of fusion penalty $\alpha$ for the effect estimates of two genes in nodes 5, 12, and 13 of FusedTree. The vertical dotted line (at $\log{\alpha}=9.6$) indicates the tuned $\alpha$ of FusedTree.}
    \label{fig:ApplicationFig}
\end{figure}

The tree fit of FusedTree, having six leaf nodes, suggests the importance
of the clinical factor covariate stage, with stage \mbox{IV} patients
having the worst outcome as expected (Figure \ref{fig:ApplicationFig}a).
The tree incorporates interactions between stage and the molecular
clustering covariate CMS and between stage and age. CMS only interacts with stage \mbox{I} and \mbox{II} patients, as reported previously \citep{CMSxStage}. Clinical covariates
gender and the site of the tumor are not part of FusedTree.

FusedTree with omics effects in nodes $7,8,$ and $9$ removed outperforms
FusedTree without omics effect removal for all $p_{\textrm{sel}}$
(Table \ref{tab:CIndex_CRCApplication}). Removing omics effects
in more nodes degrades performance. This finding suggests that the overall omics effect is not required for prognosis for patients that 1) have a tumor in stage \mbox{I} or \mbox{II} and belong
to molecular cluster CMS1, CMS2, or CMS3 and 2) have a stage
\mbox{IV} tumor.
For patients that 1) have a tumor in stage \mbox{I} or \mbox{II} and
belong to molecular cluster CMS4 and 2) have a stage \mbox{III}
tumor, the overall omics effect improves prognosis. Apparently, the subgroups
with the best prognosis (most left two nodes of the tree) and the
poorest prognosis (most right node of the tree) do not require omics
effects.

FusedTree (with omics effect removal) tunes $\lambda=1508$ and fusion penalty $\alpha=14836$
Figure \ref{fig:ApplicationFig}b shows regularization paths of the effect sizes of genes MAGEA6 and HLA-DRB4 as a function of $\alpha$ at the tuned $\lambda$ (vertical dotted line indicates the tuned $\alpha$). These two genes show the greatest variability across the leaf nodes. Figure \ref{fig:ApplicationFig}b reveals that, at $\alpha=14836,$ interaction effects between clinical and omics covariates are present but that these effects are substantially shrunken.

\begin{table}
\centering{}\caption{Concordance index (C-index)
and time-dependent AUC (with $5$ years cut-off) of CRC prognosis of several survival
models. The performance measures are estimated on an independent test
set with $N_{\textrm{test}}=167.$ Because of memory issues, results for gradient boosting with $p_{\textrm{sel}}=21,292$ are missing.}
\begin{tabular}{lcccccc}
 & \multicolumn{2}{c}{$\boldsymbol{p_{\textrm{sel}}=500}$} & \multicolumn{2}{c}{$\boldsymbol{p_{\textrm{sel}}=5000}$} & \multicolumn{2}{c}{$\boldsymbol{p_{\textrm{sel}}=21,292}$}\tabularnewline
 &C-index&t-AUC&C-index&t-AUC&C-index&t-AUC\tabularnewline
\hline 
\hline 
FusedTree & $0.72$ & $0.77$ & $0.73$ & $0.74$ & $0.73$ & $0.75$\tabularnewline
FusedTree N7,N8,N9 & $0.75$ & $0.79$ & $0.76$ & $0.77$ & $\boldsymbol{0.76}$ & $\boldsymbol{0.77}$\tabularnewline
Cox PH (clinical only) & $0.72$ & $0.69$ & $0.72$ & $0.69$ & $0.72$ & $0.69$\tabularnewline
Ridge & $0.73$ & $0.73$ & $0.73$ & $0.72$ & $0.73$ & $0.72$\tabularnewline
Lasso & $0.71$ & $0.72$ & $0.71$ & $0.71$ & $0.73$ & $0.72$\tabularnewline
Gradient Boosting & $0.69$ & $0.74$ & $0.68$ & $0.67$ & - & -\tabularnewline
Random forest & $0.71$ & $0.74$ & $0.68$ & $0.71$ & $0.62$ & $0.64$\tabularnewline
Block Forest & $\boldsymbol{0.77}$ & \textbf{$\boldsymbol{0.80}$} & $\boldsymbol{0.77}$ & $\boldsymbol{0.78}$ & $0.75$ & $0.75$\tabularnewline
\end{tabular}
\label{tab:CIndex_CRCApplication}
\end{table}

Among competitors, we first compare FusedTree (omics effect removed
in nodes $7,8,$ and $9$) with the linear models. FusedTree performs
substantially better than the clinical cox model and ridge and lasso
regression perform marginally better, which suggests that the omics
covariate set improves prognosis on top of the clinical covariate
set. The comparative performance of FusedTree
and ridge implies that FusedTree better approximates the
prognostic clinical covariate part by modeling interactions and by
more naturally handling categorical covariates. Additionally, the shrunken clinical $\times$ omics interaction effects may enhance prognosis. FusedTree and linear
competitors do not show a decline in performance for larger number
of omics covariates. 

Among nonlinear models, FusedTree is competitive to block forest, and FusedTree outperforms gradient boosting and standard random forest. We do not have results for gradient boosting for all omics covariates ($p_{\textrm{sel}}=21,292$) because we ran into memory issues. Random forest and gradient boosting show strong decline in performance for larger $p_{\textrm{sel}}$. This decline suggests that nonlinear models have difficulty in finding the prognostic signal when many (noisy) covariates are added. These models require \textit{a priori} favoring of the clinical covariate set, as indicated by the comparative performance of block forest and random forest. However, for $p_{\textrm{sel}}=21,292$, the performance of block forest also decreases. 

A strong benefit of FusedTree, in particular with respect to variations of the random forest such as block forest, is its interpretability on various levels: the relevance of the clinical covariates is easily extracted from the single tree,  whereas the regression coefficients  allow quantification of relevance of genomics for patient subgroups. We illustrate the interpretability of FusedTree for the CRC application below.

First, the fitted FusedTree model suggests that for patient subpopulations defined by leaf node 7, 8, and 9 the omics effects do not add to prognosis. Second, the regularization paths in Figure \ref{fig:ApplicationFig}b indicate that overall interactions between clinical and omics covariates in the nonzero leaf nodes (5, 12, and 13) are weak. Third, the sum of absolute omics effect size estimates is largest in leaf node 12: ($\left\Vert\boldsymbol{\beta}_{N5}\right\Vert _{1}=10.7,$ $\left\Vert\boldsymbol{\beta}_{N12}\right\Vert _{1}=11.9,$ and $\left\Vert\boldsymbol{\beta}_{N13}\right\Vert _{1}=10.1$). This finding suggests that omics covariates have the strongest overall effect on prognosis of patients younger than $80$ years with a stage \mbox{III} tumor. Fourth, the variance of gene effect size estimates across nodes is informative. For example, the MAGE-A set of genes is over-represented in the top $20$ of genes with the largest variance across nodes (e.g. Figure \ref{fig:ApplicationFig}b). This set of genes expresses cancer/testis (CT) antigens and is therefore important in immunotherapy \citep{MageACRC}. This variability may turn out valuable for e.g. heterogeneous treatment estimation because the prognostic effect of immunotherapy may vary across patient subpopulations. Last, the total absolute sum of effect size estimates of a recently published gene signature associated with CRC prognosis \citep{SignatureCRC} is twice as large in node 13 compared to node 5 and 12, suggesting a difference in importance of this signature across different subpopulations. 

\section{Conclusion}
\label{sec:Discussion}
We developed FusedTree, a model that deals with high-dimensional omics covariates and well-established clinical risk factors by combining a regression tree with fusion-like ridge regression. We showed the benefits of the fusion penalty in simulations. An application to colorectal cancer prognosis  illustrated that FusedTree 1) had a better model fit compared to several competitors and 2) rendered insights in the added overall benefit of omics measurements to prognosis for different patient subgroups compared to only employing clinical risk factors.

We opted for fitting the penalized regression conditional on the tree instead of optimizing the regression and tree jointly as is considered by \cite{ZeileisModelBasedPartitioning}. The conditional strategy puts more weight on the clinical covariates that define the tree and is therefore more consistent with the established prognostic effect of these covariates. Furthermore,  joint optimization is challenging because the omics data is high-dimensional and because optimizing a tree is a non-convex and non-smooth problem. One solution may be to embed FusedTree in a Bayesian framework by employing Bayesian CART model search \citep{BayesCART} for the tree combined with linear regressions with normal priors. This approach, however, is computationally intensive and model interpretations from the sampled tree posterior will likely be more challenging than for our current solution.

Additional structures may be incorporated into FusedTree. For example, the fusion strength may decrease with a distance measure between leaf nodes. \citet{JMLRLaplacian} proposed a related strategy in which interaction effects were weaker for more similar instances of the effect modifiers. Defining a generic distance measure for the leaf nodes of FusedTree is nontrivial because the difference in interaction strength between leaf nodes depends on the characteristic of variables employed in the splitting rules.


\section{Data availability and software}
\label{Sofware}
Data of the colorectal cancer application are publicly available in the R package \href{https://bioconductor.org/packages/release/data/experiment/html/mcsurvdata.html}{\texttt{mcsurvdata}}. These data and R code (version 4.4.1) to reproduce results presented in Section \ref{sec:Simulations}  and \ref{sec:Application} are available via \url{https://github.com/JeroenGoedhart/FusedTree_paper}.

\section*{Competing interests}
No competing interest is declared.

\section*{Acknowledgments}
The authors thank Hanarth Fonds for their financial support.

\end{document}


\sloppy
\begin{center}
\large\textbf{SUPPLEMENTARY MATERIAL TO: Fusion of Tree-induced Regressions for Clinico-genomic Data}
\end{center}\bigskip{}

\begin{center}
Jeroen M. Goedhart$^\ast$\textsuperscript{a},  Mark A. van de Wiel\textsuperscript{a}, Wessel N. van Wieringen\textsuperscript{a,b}, Thomas Klausch\textsuperscript{a}
\end{center}

\begin{center}
$^\ast$\small{Correspondence e-mail address: j.m.goedhart@amsterdamumc.nl}
\end{center}

\begin{center}
\textsuperscript{a}\small\it{Department of Epidemiology and Data Science, Amsterdam Public Health Research Institute,
Amsterdam University Medical Centers Location AMC, Meibergdreef 9,
the Netherlands}\\
\textsuperscript{b}\small\it{Department of Mathematics, Vrije Universiteit, De Boelelaan 1081a, 1081 HV Amsterdam, The Netherlands}
\end{center}

\section{FusedTree for binary outcome}
\label{sec:FusedTree-for-binary}

Recall that the fitted tree with $M$ leaf nodes induces data $\tilde{\boldsymbol{y}}\in\mathbb{R}^{N\times1},$
$\tilde{\boldsymbol{X}}\in\mathbb{R}^{N\times Mp},$ and $\tilde{\boldsymbol{U}}\in\mathbb{R}^{N\times M}.$
We index observations, corresponding to the rows, of $\tilde{\boldsymbol{y}},$
$\tilde{\boldsymbol{X},}$ and $\tilde{\boldsymbol{U}}$ by $i,$
i.e. $\tilde{y}_{i},$ $\tilde{\boldsymbol{x}}_{i},$ and $\tilde{\boldsymbol{u}}_{i}$
. Then, for binary response $y_{i}\in\left\{ 0,1\right\} ,$ we consider
the model
\begin{equation}
\tilde{y}_{i}  \sim\textrm{Bernoulli}\left[\textrm{expit}\left(\tilde{\boldsymbol{u}}_{i}^{\top}\boldsymbol{c}+\tilde{\boldsymbol{x}}_{i}^{\top}\boldsymbol{\beta}\right)\right],\quad i=1,\ldots,N,\label{eq:LogisticModel}
\end{equation}
with again clinical intercept parameter vector $\boldsymbol{c}\in\mathbb{R}^{M},$
omics parameter vector $\boldsymbol{\beta}\in\mathbb{R}^{Mp},$ and
$\textrm{expit\ensuremath{\left(x\right)}=}\exp\left(x\right)\left[1+\exp\left(x\right)\right]^{-1}.$
To find estimates $\hat{\boldsymbol{c}}$ of \textbf{$\boldsymbol{c}$}
and $\hat{\boldsymbol{\beta}}$ and $\boldsymbol{\beta},$ we solve
\begin{align}
\hat{\boldsymbol{c}},\,\hat{\textbf{\ensuremath{\boldsymbol{\beta}}}} & =\underset{\boldsymbol{c},\textbf{\ensuremath{\boldsymbol{\beta}}}}{\textrm{arg max}}\:\sum_{i=1}^{N}\tilde{y}_{i}\log\left[\textrm{expit}\left(\tilde{\boldsymbol{u}}_{i}^{\top}\boldsymbol{c}+\tilde{\boldsymbol{x}}_{i}^{\top}\boldsymbol{\beta}\right)\right]+\left(1-\tilde{y}_{i}\right)\log\left[1-\textrm{expit}\left(\tilde{\boldsymbol{u}}_{i}^{\top}\boldsymbol{c}+\tilde{\boldsymbol{x}}_{i}^{\top}\boldsymbol{\beta}\right)\right]\nonumber \\
& -\lambda\textbf{\ensuremath{\boldsymbol{\beta}}}^{T}\textbf{\ensuremath{\boldsymbol{\beta}}}-\alpha\textbf{\ensuremath{\boldsymbol{\beta}}}^{\top}\boldsymbol{\Omega}\textbf{\ensuremath{\boldsymbol{\beta}}}\nonumber \\
 & =\underset{\boldsymbol{c},\textbf{\ensuremath{\boldsymbol{\beta}}}}{\textrm{arg max}}\:\sum_{i=1}^{N}\left\{ \tilde{y}_{i}\left(\tilde{\boldsymbol{u}}_{i}^{\top}\boldsymbol{c}+\tilde{\boldsymbol{x}}_{i}^{\top}\boldsymbol{\beta}\right)-\log\left[1+\exp\left(\tilde{\boldsymbol{u}}_{i}^{\top}\boldsymbol{c}+\tilde{\boldsymbol{x}}_{i}^{\top}\boldsymbol{\beta}\right)\right]\right\} -\lambda\textbf{\ensuremath{\boldsymbol{\beta}}}^{T}\textbf{\ensuremath{\boldsymbol{\beta}}}-\alpha\textbf{\ensuremath{\boldsymbol{\beta}}}^{\top}\boldsymbol{\Omega}\textbf{\ensuremath{\boldsymbol{\beta}}},\label{eq:EstimationLogModel}
\end{align}
i.e. optimizing the penalized log likelihood of all data for model
(\ref{eq:LogisticModel}). 

Estimator (\ref{eq:EstimationLogModel}) cannot be evaluated analytically
and is hence found using the iterative re-weighted least squares (IRLS)
algorithm.2

The IRLS algorithm updates estimates $\hat{\boldsymbol{c}}^{(t)},\,\hat{\textbf{\ensuremath{\boldsymbol{\beta}}}}^{(t)}\rightarrow\hat{\boldsymbol{c}}^{(t+1)},\,\hat{\textbf{\ensuremath{\boldsymbol{\beta}}}}^{(t+1)},$
with iteration index $t,$ until the estimates stabilize within some
tolerance level. Specifically, define the linear predictor for the
observations $\boldsymbol{\eta}^{(t)}=\left(\eta_{i}^{(t)}\right)_{i=1}^{N}\in\mathbb{R}^{N\times1},$
with $\eta_{i}^{(t)}=\tilde{\boldsymbol{u}}_{i}^{\top}\hat{\boldsymbol{c}}^{(t)}+\tilde{\boldsymbol{x}}_{i}^{\top}\hat{\textbf{\ensuremath{\boldsymbol{\beta}}}}^{(t)}$,
diagonal weight matrix $\boldsymbol{W}^{(t)}$ with $i$th element
on the diagonal $W_{ii}^{(t)}=\exp\left(\eta_{i}^{(t)}\right)\left[\exp\left(\eta_{i}^{(t)}\right)+1\right]^{-2},$
Then, given current estimates $\hat{\boldsymbol{c}}^{(t)},\,\hat{\textbf{\ensuremath{\boldsymbol{\beta}}}}^{(t)},$
the updates equal: 
\begin{align}
\hat{\boldsymbol{c}}^{(t+1)} & =\left\{ \tilde{\boldsymbol{U}}^{\top}\left(\boldsymbol{W}^{(t)}\right)^{-1}\left[\textbf{\ensuremath{\tilde{\boldsymbol{X}}}}\left(\lambda\boldsymbol{I}_{Mp\times Mp}+\alpha\boldsymbol{\Omega}\right)^{-1}\textbf{\ensuremath{\tilde{\boldsymbol{X}}}}^{\top}+\left(\boldsymbol{W}^{(t)}\right)^{-1}\right]^{-1}\tilde{\boldsymbol{U}}\right\} ^{-1}\nonumber \\
 & \times\tilde{\boldsymbol{U}}^{\top}\left[\textbf{\ensuremath{\tilde{\boldsymbol{X}}}}\left(\lambda\boldsymbol{I}_{Mp\times Mp}+\alpha\boldsymbol{\Omega}\right)^{-1}\textbf{\ensuremath{\tilde{\boldsymbol{X}}}}^{\top}+\left(\boldsymbol{W}^{(t)}\right)^{-1}\right]^{-1}\left\{ \boldsymbol{\eta}^{(t)}+\left(\boldsymbol{W}^{(t)}\right)^{-1}\left[\tilde{\boldsymbol{y}}-\textrm{expit}\left(\boldsymbol{\eta}^{(t)}\right)\right]\right\} \nonumber \\
\hat{\textbf{\ensuremath{\boldsymbol{\beta}}}}^{(t+1)} & =\left(\textbf{\ensuremath{\tilde{\boldsymbol{X}}}}^{\top}\boldsymbol{W}^{(t)}\textbf{\ensuremath{\tilde{\boldsymbol{X}}}}+\lambda\boldsymbol{I}_{Mp\times Mp}+\alpha\boldsymbol{\Omega}\right)^{-1}\textbf{\ensuremath{\tilde{\boldsymbol{X}}}}^{\top}\left[\boldsymbol{W}^{(t)}\left(\boldsymbol{\eta}^{(t)}-\tilde{\boldsymbol{U}}\hat{\boldsymbol{c}}^{(t+1)}\right)+\tilde{\boldsymbol{y}}-\textrm{expit}\left(\boldsymbol{\eta}^{(t)}\right)\right],\label{eq:LogModelEstimators}
\end{align}
as was shown by \citet{10.1093/jrsssc/qlad041}. We run the iterative
algorithm until the penalized likelihood has stabilized within an
absolute tolerance $\textrm{tol}=10^{-10}.$ 

\section{FusedTree for survival outcome}
\label{sec:FusedTree-for-survival}

For survival data, we have response $\tilde{y}_{i}=\left(t_{i},\delta_{i}\right)$
for observations $i=1,\ldots,N$ with $t_{i}$ the observed time at
which patients had an event $\left(\delta_{i}=1\right)$ or were censored
$\left(\delta_{i}=0\right).$ Again, we have tree-induced data $\tilde{\boldsymbol{X}}\in\mathbb{R}^{N\times Mp},$
and $\tilde{\boldsymbol{U}}\in\mathbb{R}^{N\times M}.$ We impose
a proportional hazards model $h\left(t\mid X_{i}\right)=h_{0}\left(t\right)\exp\left(\tilde{\boldsymbol{u}}_{i}^{\top}\boldsymbol{c}+\tilde{\boldsymbol{x}}_{i}\boldsymbol{\beta}\right),$
which induces the penalized full log likelihood
\begin{align}
l^{pen}\left(\boldsymbol{\beta},\boldsymbol{c},h_{0}\left(t\right);\tilde{\boldsymbol{y}},\tilde{\boldsymbol{X},}\tilde{\boldsymbol{U}}\right)& =\sum_{i=1}^{N}\left\{ -\exp\left(\tilde{\boldsymbol{u}}_{i}^{\top}\boldsymbol{c}+\tilde{\boldsymbol{x}}_{i}^{\top}\boldsymbol{\beta}\right)H_{0}(t_{i})+\delta_{i}\left[\log\left(h_{0}(t_{i})\right)+\tilde{\boldsymbol{u}}_{i}^{\top}\boldsymbol{c}+\tilde{\boldsymbol{x}}_{i}^{\top}\boldsymbol{\beta}\right]\right\}\nonumber \\
& - \lambda\textbf{\ensuremath{\boldsymbol{\beta}}}^{T}\textbf{\ensuremath{\boldsymbol{\beta}}}-\alpha\textbf{\ensuremath{\boldsymbol{\beta}}}^{\top}\boldsymbol{\Omega}\textbf{\ensuremath{\boldsymbol{\beta}}},\label{eq:PenLikSurvival}
\end{align}

with baseline hazard $h_{0}\left(t\right)$ and cumulative baseline
hazard $H_{0}\left(t\right)=\int_{t'=0}^{t}h_{0}\left(t'\right)dt'.$
We then aim to find estimators $\hat{\boldsymbol{c}},\,\hat{\textbf{\ensuremath{\boldsymbol{\beta}}}}$
by 
\begin{equation}
\hat{\boldsymbol{c}},\,\hat{\textbf{\ensuremath{\boldsymbol{\beta}}}}=\underset{\boldsymbol{c},\textbf{\ensuremath{\boldsymbol{\beta}}}}{\textrm{arg max}}\:l^{pen}\left(\boldsymbol{\beta},\boldsymbol{c},h_{0}\left(t\right);\tilde{\boldsymbol{y}},\tilde{\boldsymbol{X},}\tilde{\boldsymbol{U}}\right).\label{eq:EstimatorsSurv}
\end{equation}
To solve (\ref{eq:EstimatorsSurv}), we use the iterative re-weighted
least squares (IRLS) algorithm proposed by \citet{HouwelingenCox}.
Conveniently, this algorithm is almost identical to the IRLS algorithm
for logistic regression, i.e. (\ref{eq:LogModelEstimators}), as shown
by \citet{Multiridge}. The only differences between logistic regression
and penalized cox regression are weights $W_{ii}^{(t)},$ which for
penalized cox regression become $W_{ii}^{(t)}=\hat{H}_{0}^{(t)}\left(t\right)\exp\left(\tilde{\boldsymbol{u}}_{i}^{\top}\hat{\boldsymbol{c}}^{(t)}+\tilde{\boldsymbol{x}}_{i}^{\top}\hat{\textbf{\ensuremath{\boldsymbol{\beta}}}}^{(t)}\right),$
and centered response $\tilde{\boldsymbol{y}}-\textrm{expit}\left(\boldsymbol{\eta}^{(t)}\right),$
which equals $\tilde{\boldsymbol{y}}-\textrm{diag}\left(\boldsymbol{W}^{(t)}\right)$
for penalized cox regression. These changes are plugged into (\ref{eq:LogModelEstimators})
and we run the iterative algorithm until penalized likelihood (\ref{eq:PenLikSurvival})
has stabilized within an absolute tolerance $\textrm{tol}=10^{-10}$.

For iterative estimates $\hat{H}_{0}^{(t)}\left(t\right)$ of baseline
hazard $H_{0}\left(t\right),$ we employ the Breslow estimator: $\hat{H}_{0}^{(t)}\left(t\right)=\sum_{i:\,t_{i}\leq t}\left\{ \delta_{i}\left[\sum_{j:\,t_{j}\geq t_{i}}\exp\left(\tilde{\boldsymbol{u}}_{i}^{\top}\hat{\boldsymbol{c}}^{(t-1)}+\tilde{\boldsymbol{x}}_{i}^{\top}\hat{\textbf{\ensuremath{\boldsymbol{\beta}}}}^{(t-1)}\right)\right]^{-1}\right\} .$

\section{Hyper-parameter tuning}
\label{sec:Hyperparameter-tuning}

To tune hyperparameters $\alpha$ and $\lambda,$ we solve for continuous
response:
\begin{equation}
\hat{\lambda},\,\hat{\alpha}=\underset{\lambda,\,\alpha}{\textrm{arg min}}\frac{1}{K}\sum_{k=1}^{K}\left\Vert \tilde{\boldsymbol{y}}_{\varGamma_{k}}-\tilde{\boldsymbol{U}}_{\varGamma_{k}}\hat{\boldsymbol{c}}_{-\varGamma_{k}}\left(\lambda,\alpha\right)-\textbf{\ensuremath{\tilde{\boldsymbol{X}}}}_{\varGamma_{k}}\textbf{\ensuremath{\hat{\boldsymbol{\beta}}}}_{-\varGamma_{k}}\left(\lambda,\alpha\right)\right\Vert _{2}^{2},\quad\textrm{subject to}\quad\lambda,\,\alpha>0,\label{eq:HypTune_linear}
\end{equation}
and for binary response, we solve:
\begin{align}
\hat{\lambda},\,\hat{\alpha} & =\underset{\lambda,\,\alpha}{\textrm{arg min}}\frac{1}{K}\sum_{k=1}^{K}\left\{ \sum_{i\in\varGamma_{k}}\tilde{y}_{i}\left(\tilde{\boldsymbol{u}}_{i}\hat{\boldsymbol{c}}_{-\varGamma_{k}}\left(\lambda,\alpha\right)+\tilde{\boldsymbol{x}}_{i}^{\top}\textbf{\ensuremath{\hat{\boldsymbol{\beta}}}}_{-\varGamma_{k}}\left(\lambda,\alpha\right)\right)\right\} \nonumber\\
& -\frac{1}{K}\sum_{k=1}^{K}\left\{ \sum_{i\in\varGamma_{k}}\log\left[1+\exp\left(\tilde{\boldsymbol{u}}_{i}^{\top}\hat{\boldsymbol{c}}_{-\varGamma_{k}}\left(\lambda,\alpha\right)+\tilde{\boldsymbol{x}}_{i}^{\top}\textbf{\ensuremath{\hat{\boldsymbol{\beta}}}}_{-\varGamma_{k}}\left(\lambda,\alpha\right)\right)\right]\right\} \nonumber \\
& \quad\textrm{subject to}\quad\lambda,\,\alpha>0,\label{eq:HypTune_logistic} 
\end{align}

and for survival response, we solve:
\begin{align}
\hat{\lambda},\,\hat{\alpha}=\underset{\lambda,\,\alpha}{\textrm{arg min}} & \frac{1}{K}\sum_{k=1}^{K}\left\{ \sum_{i\in\varGamma_{k}}-\exp\left(\tilde{\boldsymbol{u}}_{i}^{\top}\hat{\boldsymbol{c}}_{-\varGamma_{k}}\left(\lambda,\alpha\right)+\tilde{\boldsymbol{x}}_{i}^{\top}\textbf{\ensuremath{\hat{\boldsymbol{\beta}}}}_{-\varGamma_{k}}\left(\lambda,\alpha\right)\right)\hat{H}_{0}(t_{i})\right\} \nonumber \\
 & +\frac{1}{K}\sum_{k=1}^{K}\left\{ \sum_{i\in\varGamma_{k}}\delta_{i}\left[\log\left(\hat{h}{}_{0}(t_{i})\right)+\tilde{\boldsymbol{u}}_{i}^{\top}\hat{\boldsymbol{c}}_{-\varGamma_{k}}\left(\lambda,\alpha\right)+\tilde{\boldsymbol{x}}_{i}^{\top}\textbf{\ensuremath{\hat{\boldsymbol{\beta}}}}_{-\varGamma_{k}}\left(\lambda,\alpha\right)\right]\right\} \\
 & \quad\textrm{subject to}\quad\lambda,\,\alpha>0,\nonumber 
\end{align}
with $\varGamma_{k}$ the observations in test fold $K$ and $-\varGamma_{k}$
the remaining samples forming the training set. Thus, we select $\hat{\lambda},\,\hat{\alpha}$
by minimizing the cross-validated prediction mean square error for
continuous $\tilde{y}_{i}$ and the cross-validated likelihood for
binary and survival $\tilde{y}_{i}.$

The above optimizations depend on repeated evaluation of estimators
$\hat{\boldsymbol{c}}_{-\varGamma_{k}}\left(\lambda,\alpha\right)$
and $\textbf{\ensuremath{\hat{\boldsymbol{\beta}}}}_{-\varGamma_{k}}\left(\lambda,\alpha\right),$
which requires considerable computational time for high-dimensional
data. As was shown by \citet{Multiridge}, a computationally more efficient
procedure is to directly evaluate the linear predictors $\tilde{\boldsymbol{U}}_{\varGamma_{k}}\hat{\boldsymbol{c}}_{-\varGamma_{k}}\left(\lambda,\alpha\right)$
and $\textbf{\ensuremath{\tilde{\boldsymbol{X}}}}_{\varGamma_{k}}\textbf{\ensuremath{\hat{\boldsymbol{\beta}}}}_{-\varGamma_{k}}\left(\lambda,\alpha\right),$
i.e. the estimators in combination with their corresponding design
matrices. These linear predictors can be reformulated such that their
evaluation only requires repeated operations on matrices of dimension
$N-\left|\varGamma_{k}\right|$ instead of dimension $Mp$ for evaluation
of $\hat{\boldsymbol{c}}_{-\varGamma_{k}}\left(\lambda,\alpha\right)$
and $\textbf{\ensuremath{\hat{\boldsymbol{\beta}}}}_{-\varGamma_{k}}\left(\lambda,\alpha\right).$
The linear predictors are given, as derived by \citet{10.1093/jrsssc/qlad041},
with $\check{\boldsymbol{X}}=\tilde{\boldsymbol{X}}\boldsymbol{V}_{\varOmega}\left(\lambda\boldsymbol{I}_{p\times p}+\alpha\boldsymbol{D}_{\varOmega}\right)^{-\frac{1}{2}},$
by 
\begin{align*}
\tilde{\boldsymbol{U}}_{\varGamma_{k}}\hat{\boldsymbol{c}}_{-\varGamma_{k}}\left(\lambda,\alpha\right) & =\tilde{\boldsymbol{U}}_{\varGamma_{k}}\left[\tilde{\boldsymbol{U}}_{-\varGamma_{k}}^{\top}\left(\check{\boldsymbol{X}}_{-\varGamma_{k}}\check{\boldsymbol{X}}_{-\varGamma_{k}}^{\top}+\boldsymbol{I}_{\left|-\varGamma_{k}\right|\times\left|-\varGamma_{k}\right|}\right)^{-1}\tilde{\boldsymbol{U}}_{-\varGamma_{k}}\right]^{-1}\\
 & \times\tilde{\boldsymbol{U}}_{-\varGamma_{k}}^{\top}\left(\check{\boldsymbol{X}}_{-\varGamma_{k}}\check{\boldsymbol{X}}_{-\varGamma_{k}}^{\top}+\boldsymbol{I}_{\left|-\varGamma_{k}\right|\times\left|-\varGamma_{k}\right|}\right)^{-1}\tilde{\boldsymbol{y}}_{-\varGamma_{k}}\\
\textbf{\ensuremath{\tilde{\boldsymbol{X}}}}_{\varGamma_{k}}\textbf{\ensuremath{\hat{\boldsymbol{\beta}}}}_{-\varGamma_{k}}\left(\lambda,\alpha\right) & =\check{\boldsymbol{X}}_{\varGamma_{k}}\check{\boldsymbol{X}}_{-\varGamma_{k}}^{\top}\left(\check{\boldsymbol{X}}_{-\varGamma_{k}}\check{\boldsymbol{X}}_{-\varGamma_{k}}^{\top}+\boldsymbol{I}_{\left|-\varGamma_{k}\right|\times\left|-\varGamma_{k}\right|}\right)^{-1}\left(\tilde{\boldsymbol{y}}_{-\varGamma_{k}}-\tilde{\boldsymbol{U}}_{-\varGamma_{k}}\hat{\boldsymbol{c}}_{-\varGamma_{k}}\right),
\end{align*}
for continuous response, and 
\begin{align*}
\tilde{\boldsymbol{U}}_{\varGamma_{k}}\hat{\boldsymbol{c}}_{-\varGamma_{k}}^{(t+1)}\left(\lambda,\alpha\right) & =\tilde{\boldsymbol{U}}_{\varGamma_{k}}\left\{\tilde{\boldsymbol{U}}_{-\varGamma_{k}}^{\top}\left(\boldsymbol{W}_{-\varGamma_{k},-\varGamma_{k}}^{(t)}\right)^{-1}\left[\check{\boldsymbol{X}}_{-\varGamma_{k}}\check{\boldsymbol{X}}_{-\varGamma_{k}}^{\top}+\left(\boldsymbol{W}_{-\varGamma_{k},-\varGamma_{k}}^{(t)}\right)^{-1}\right]^{-1}\tilde{\boldsymbol{U}}_{-\varGamma_{k}}\right\} ^{-1}\\
 & \times\tilde{\boldsymbol{U}}_{-\varGamma_{k}}^{\top}\left[\check{\boldsymbol{X}}_{-\varGamma_{k}}\check{\boldsymbol{X}}_{-\varGamma_{k}}^{\top}+\left(\boldsymbol{W}_{-\varGamma_{k},-\varGamma_{k}}^{(t)}\right)^{-1}\right]^{-1}\left\{ \boldsymbol{\eta}_{-\varGamma_{k}}^{(t)}+\left(\boldsymbol{W}_{-\varGamma_{k},-\varGamma_{k}}^{(t)}\right)^{-1}\right.\\
 & \left.\times
 \left[\tilde{\boldsymbol{y}}_{-\varGamma_{k}}-\textrm{expit}\left(\boldsymbol{\eta}_{-\varGamma_{k}}^{(t)}\right)\right]\right\} \\
\textbf{\ensuremath{\tilde{\boldsymbol{X}}}}_{\varGamma_{k}}\textbf{\ensuremath{\hat{\boldsymbol{\beta}}}}_{-\varGamma_{k}}^{(t+1)}\left(\lambda,\alpha\right) & =\check{\boldsymbol{X}}_{\varGamma_{k}}\check{\boldsymbol{X}}_{-\varGamma_{k}}^{\top}\left[\check{\boldsymbol{X}}_{-\varGamma_{k}}\check{\boldsymbol{X}}_{-\varGamma_{k}}^{\top}+\left(\boldsymbol{W}_{-\varGamma_{k},-\varGamma_{k}}^{(t)}\right)^{-1}\right]^{-1}\\
& \times\left\{\boldsymbol{\eta}_{-\varGamma_{k}}^{(t)}-\tilde{\boldsymbol{U}}_{-\varGamma_{k}}\hat{\boldsymbol{c}}_{-\varGamma_{k}}^{(t+1)}+\left(\boldsymbol{W}_{-\varGamma_{k},-\varGamma_{k}}^{(t)}\right)^{-1}\left[\tilde{\boldsymbol{y}}_{-\varGamma_{k}}-\textrm{expit}\left(\boldsymbol{\eta}_{-\varGamma_{k}}^{(t)}\right)\right]\right\},
\end{align*}
for binary response, with diagonal weight matrix $W_{-\varGamma_{k},-\varGamma_{k}}^{(t)}$
and linear predictor $\boldsymbol{\eta}_{-\varGamma_{k}}^{(t)}$ defined
as in Appendix \ref{sec:FusedTree-for-binary} combined with appropriate
subsetting. Again, for survival response, we use a similar algorithm
as for binary response in which only weights $W_{-\varGamma_{k},-\varGamma_{k}}^{(t)}$
and $\tilde{\boldsymbol{y}}_{-\varGamma_{k}}-\textrm{expit}\left(\boldsymbol{\eta}_{-\varGamma_{k}}^{(t)}\right)$
are modified as described in Appendix \ref{sec:FusedTree-for-survival}
.

Optimizations \ref{eq:HypTune_linear} and \ref{eq:HypTune_logistic}
are performed using the Nelder-Mead method (\citealp{NelderMead}) implemented
in the base R \texttt{optim} function with penalties $\lambda,\alpha$
on the $\log$-scale.

\section{Shrinkage limits}
\label{sec:Shrinkage-limits}

Here, we derive the shrinkage limits of the FusedTree estimator, which
we presented in eq. 6 of the main text.

Define $\boldsymbol{\varLambda}_{\lambda,\alpha}=\lambda\boldsymbol{I}_{Mp\times Mp}+\alpha\boldsymbol{\Omega},$
and recall $\boldsymbol{\Omega}=\boldsymbol{I}_{p\times p}\bigotimes\left(\boldsymbol{I}_{M\times M}-\frac{1}{M}\boldsymbol{1}_{M\times M}\right),$
with $M$ the number of leaf nodes. The estimators for the tree-induced
clinical effect $\boldsymbol{c}$ and omics effects $\boldsymbol{\beta}$
are
\begin{align}
\hat{\boldsymbol{c}} & =\left\{ \tilde{\boldsymbol{U}}^{\top}\left[\textbf{\ensuremath{\tilde{\boldsymbol{X}}}}\boldsymbol{\varLambda}_{\lambda,\alpha}^{-1}\textbf{\ensuremath{\tilde{\boldsymbol{X}}}}^{\top}+\boldsymbol{I}_{N\times N}\right]^{-1}\tilde{\boldsymbol{U}}\right\} ^{-1}\nonumber \\
 & \times\tilde{\boldsymbol{U}}^{\top}\left[\textbf{\ensuremath{\tilde{\boldsymbol{X}}}}\boldsymbol{\varLambda}_{\lambda,\alpha}^{-1}\textbf{\ensuremath{\tilde{\boldsymbol{X}}}}^{\top}+\boldsymbol{I}_{N\times N}\right]^{-1}\tilde{\boldsymbol{y}}\nonumber \\
\hat{\textbf{\ensuremath{\boldsymbol{\beta}}}} & =\left(\textbf{\ensuremath{\tilde{\boldsymbol{X}}}}^{\top}\textbf{\ensuremath{\tilde{\boldsymbol{X}}}}+\boldsymbol{\varLambda}_{\lambda,\alpha}\right)^{-1}\textbf{\ensuremath{\tilde{\boldsymbol{X}}}}^{\top}\left(\tilde{\boldsymbol{y}}-\tilde{\boldsymbol{U}}\hat{\boldsymbol{c}}\right),\nonumber \\
 & =\left[\boldsymbol{\varLambda}_{\lambda,\alpha}^{-1}-\boldsymbol{\varLambda}_{\lambda,\alpha}^{-1}\textbf{\ensuremath{\tilde{\boldsymbol{X}}}}^{\top}\left(\textbf{\ensuremath{\tilde{\boldsymbol{X}}}}\boldsymbol{\varLambda}_{\lambda,\alpha}^{-1}\textbf{\ensuremath{\tilde{\boldsymbol{X}}}}^{\top}+\boldsymbol{I}_{N\times N}\right)^{-1}\textbf{\ensuremath{\tilde{\boldsymbol{X}}}}\boldsymbol{\varLambda}_{\lambda,\alpha}^{-1}\right]\textbf{\ensuremath{\tilde{\boldsymbol{X}}}}^{\top}\left(\tilde{\boldsymbol{y}}-\tilde{\boldsymbol{U}}\hat{\boldsymbol{c}}\right)\label{eq:EstimatorsAppendix}
\end{align}
with the last line of (\ref{eq:EstimatorsAppendix}) following from
Woodbury's identity. To derive the shrinkage limits ($\lambda\rightarrow\infty$
and $\alpha\rightarrow\infty$) of (\ref{eq:EstimatorsAppendix}),
we first find $\boldsymbol{\varLambda}_{\lambda,\alpha}^{-1}.$ Because
$\boldsymbol{\varLambda}_{\lambda,\alpha}=\boldsymbol{I}_{p\times p}\bigotimes\boldsymbol{A},$
with $A=(\lambda+\alpha)\boldsymbol{I}_{M\times M}-\frac{\alpha}{M}\boldsymbol{1}_{M\times M},$
we have $\boldsymbol{\varLambda}_{\lambda,\alpha}^{-1}=\boldsymbol{I}_{p\times p}\bigotimes\boldsymbol{A}^{-1},$
and we are left with determining $\boldsymbol{A}^{-1},$ which can
be shown to equal
\[
\boldsymbol{A}^{-1}=\begin{pmatrix}a & b & \cdots & b\\
b & \ddots & \ddots & \vdots\\
\vdots & \ddots & a & b\\
b & \cdots & b & a
\end{pmatrix}\in\mathbb{R}^{M\times M},
\]
having identical diagonal elements $a=\lambda^{-1}-\alpha\left(1-1/M\right)\left(\lambda^{2}+\lambda\alpha\right)^{-1}$
and identical off-diagonal elements $b=\alpha\left(\lambda^{2}M+\lambda\alpha M\right)^{-1}.$
For $\lambda\rightarrow\infty,$ we have $a=b=0,$ and for $\alpha\rightarrow\infty,$
we have $a=b=1/\left(\lambda M\right).$ Thus, we have 
\begin{align}
\lim_{\lambda\rightarrow\infty}\boldsymbol{\varLambda}_{\lambda,\alpha}^{-1} & =\boldsymbol{0}_{Mp\times Mp}\label{eq:LambdaLimit}\\
\lim_{\alpha\rightarrow\infty}\boldsymbol{\varLambda}_{\lambda,\alpha}^{-1} & =\frac{1}{\lambda M}\boldsymbol{I}_{p\times p}\bigotimes\boldsymbol{1}_{M\times M}.\label{eq:AlphaLimit}
\end{align}
Limit (\ref{eq:LambdaLimit}) renders estimators$\boldsymbol{:}$
\begin{align}
\lim_{\lambda\rightarrow\infty}\hat{\boldsymbol{c}} & =\left(\tilde{\boldsymbol{U}}^{\top}\tilde{\boldsymbol{U}}\right)^{-1}\tilde{\boldsymbol{U}}^{\top}\tilde{\boldsymbol{y}}\label{eq:LamLimit}\\
\lim_{\lambda\rightarrow\infty}\hat{\textbf{\ensuremath{\boldsymbol{\beta}}}} & =\boldsymbol{0}_{Mp},\nonumber 
\end{align}
with the first line the standard normal equation, as expected.

For $\alpha\rightarrow\infty,$ we first define the face-splitting
product (\citealp{Slyusar1999}) by $\bullet,$ with matrix $\boldsymbol{C}=\boldsymbol{A}\bullet\boldsymbol{B}$
having row $i$ defined by the Kronecker product of corresponding
rows $i$ of \textbf{$\boldsymbol{A}$} and $\boldsymbol{B}.$ For
$\boldsymbol{A}\in\mathbb{R}^{N\times M}$ and $\boldsymbol{B}\in\mathbb{R}^{N\times p},$
we then have $\boldsymbol{C}\in\mathbb{R}^{N\times Mp}.$ We also
define the column-wise Kronecker product, i.e. the the Khatri--Rao
product (\citealp{KhatriRhao}), by $\ast,$ with $\boldsymbol{C}=\boldsymbol{A}\ast\boldsymbol{B}$
having column $j$ defined by the Kronecker product of column $j$
of $\boldsymbol{A}$ and $\boldsymbol{B}.$ For these products, the
following useful properties hold (\citealp{Slyusar1999}):
\begin{align*}
\left(\boldsymbol{A}\bullet\boldsymbol{B}\right)\left(\boldsymbol{C}\bigotimes\boldsymbol{D}\right) & =\left(\boldsymbol{AC}\right)\bullet\left(\boldsymbol{BD}\right)\\
\left(\boldsymbol{A}\bigotimes\boldsymbol{B}\right)\left(\boldsymbol{C}\ast\boldsymbol{D}\right) & =\left(\boldsymbol{AC}\right)\ast\left(\boldsymbol{BD}\right)\\
\left(\boldsymbol{A}\bullet\boldsymbol{B}\right)\left(\boldsymbol{C}\ast\boldsymbol{D}\right) & =\left(\boldsymbol{AC}\right)\circ\left(\boldsymbol{BD}\right)\\
\left(\boldsymbol{A}\bullet\boldsymbol{B}\right)^{\top} & =\boldsymbol{A}^{\top}\ast\boldsymbol{B}^{\top}
\end{align*}
with $\circ$ the Hadamard product, and all matrices of the right
dimension to perform multiplication. These definitions are useful
because we may define the tree-induced omics matrix $\tilde{\boldsymbol{X}}$
by 
\begin{equation}
\tilde{\boldsymbol{X}}=\boldsymbol{X}\bullet\tilde{\boldsymbol{U}},\label{eq:FaceSplitProd}
\end{equation}
with \textbf{$\boldsymbol{X}\in\mathbb{R}^{N\times p}$ }the original
omics covariate matrix.

We start with $\underset{\alpha\rightarrow\infty}{\lim}\hat{\boldsymbol{c}}.$
The limit $\underset{\alpha\rightarrow\infty}{\lim}\left[\textbf{\ensuremath{\tilde{\boldsymbol{X}}}}\boldsymbol{\varLambda}_{\lambda,\alpha}^{-1}\textbf{\ensuremath{\tilde{\boldsymbol{X}}}}^{\top}+\boldsymbol{I}_{N\times N}\right]^{-1}$
in (\ref{eq:EstimatorsAppendix}) is simplified using (\ref{eq:FaceSplitProd})
to
\begin{align}
\lim_{\alpha\rightarrow\infty}\textbf{\ensuremath{\tilde{\boldsymbol{X}}}}\boldsymbol{\varLambda}_{\lambda,\alpha}^{-1}\textbf{\ensuremath{\tilde{\boldsymbol{X}}}}^{\top} & =\frac{1}{\lambda M}\textbf{\ensuremath{\left(\boldsymbol{X}\bullet\tilde{\boldsymbol{U}}\right)}}\left(\boldsymbol{I}_{p\times p}\bigotimes\boldsymbol{1}_{M\times M}\right)\textbf{\ensuremath{\left(\boldsymbol{X}\bullet\tilde{\boldsymbol{U}}\right)}}^{\top}\nonumber \\
 & =\frac{1}{\lambda M}\textbf{\ensuremath{\left(\boldsymbol{X}\bullet\tilde{\boldsymbol{U}}\boldsymbol{1}_{M\times M}\right)}}\textbf{\ensuremath{\left(\boldsymbol{X}^{\top}\ast\tilde{\boldsymbol{U}}^{\top}\right)}}\nonumber \\
 & =\frac{1}{\lambda M}\textbf{\ensuremath{\left(\boldsymbol{X}\boldsymbol{X}^{\top}\right)}}\circ\left(\tilde{\boldsymbol{U}}\boldsymbol{1}_{M\times M}\tilde{\boldsymbol{U}}^{\top}\right)=\frac{1}{\lambda M}\textbf{\ensuremath{\boldsymbol{X}\boldsymbol{X}^{\top}}},\label{eq:UsefulIdent1}
\end{align}
where we used $\tilde{\boldsymbol{U}}\boldsymbol{1}_{M\times M}\tilde{\boldsymbol{U}}^{\top}=\boldsymbol{1}_{N\times N}.$
This leads to the following limit 
\begin{align}
\lim_{\alpha\rightarrow\infty}\hat{\boldsymbol{c}} & =\left\{ \tilde{\boldsymbol{U}}^{\top}\left[\boldsymbol{X}\left(\frac{1}{\lambda M}\boldsymbol{I}_{p\times p}\right)\boldsymbol{X}^{\top}+\boldsymbol{I}_{N\times N}\right]^{-1}\tilde{\boldsymbol{U}}\right\} ^{-1}\nonumber \\
 & \times\tilde{\boldsymbol{U}}^{\top}\left[\boldsymbol{X}\left(\frac{1}{\lambda M}\boldsymbol{I}_{p\times p}\right)\boldsymbol{X}^{\top}+\boldsymbol{I}_{N\times N}\right]^{-1}\tilde{\boldsymbol{y}}.\label{eq:AlphaLimitClinical}
\end{align}
Equation (\ref{eq:AlphaLimitClinical}) is almost identical to the
unpenalized effect estimator of a standard ridge regression with unpenalized
$\tilde{\boldsymbol{U}}$ and penalized $\boldsymbol{X}$ (so the
limit $\underset{\alpha\rightarrow\infty}{\lim}$ reduces $\textbf{\ensuremath{\tilde{\boldsymbol{X}}}}$
to $\boldsymbol{X}$). The standard ridge penalty, however, is multiplied
by $M$ in (\ref{eq:AlphaLimitClinical}) to account for having a
factor $M$ more omics effect estimates.

Next, we compute $\underset{\alpha\rightarrow\infty}{\lim}\hat{\boldsymbol{\beta}}.$
We first note the equality 
\begin{equation}
\lim_{\alpha\rightarrow\infty}\boldsymbol{\varLambda}_{\lambda,\alpha}^{-1}\textbf{\ensuremath{\tilde{\boldsymbol{X}}}}^{\top}=\frac{1}{\lambda M}\left(\boldsymbol{I}_{p\times p}\bigotimes\boldsymbol{1}_{M\times M}\right)\textbf{\ensuremath{\left(\boldsymbol{X}^{\top}\ast\tilde{\boldsymbol{U}}^{\top}\right)}}=\frac{1}{\lambda M}\boldsymbol{X}^{\top}\ast\boldsymbol{1}_{M\times N}.\label{eq:UsefulIdent2}
\end{equation}
Then, plugging (\ref{eq:UsefulIdent1}) and (\ref{eq:UsefulIdent2})
into the last line of (\ref{eq:EstimatorsAppendix}) renders 
\begin{align*}
\underset{\alpha\rightarrow\infty}{\lim}\hat{\boldsymbol{\beta}} & =\left[\frac{1}{\lambda M}\boldsymbol{X}^{\top}\ast\boldsymbol{1}_{M\times N}-\frac{1}{\lambda M}\boldsymbol{X}^{\top}\ast\boldsymbol{1}_{M\times N}\left(\boldsymbol{I}_{N\times N}+\frac{1}{\lambda M}\textbf{\ensuremath{\boldsymbol{X}\boldsymbol{X}^{\top}}}\right)^{-1}\frac{1}{\lambda M}\textbf{\ensuremath{\boldsymbol{X}\boldsymbol{X}^{\top}}}\right]\left(\tilde{\boldsymbol{y}}-\tilde{\boldsymbol{U}}\hat{\boldsymbol{c}}\right)\\
 & =\left[\frac{1}{\lambda M}\boldsymbol{X}^{\top}-\frac{1}{\left(\lambda M\right)^{2}}\boldsymbol{X}^{\top}\left(\boldsymbol{I}_{N\times N}+\frac{1}{\lambda M}\textbf{\ensuremath{\boldsymbol{X}\boldsymbol{X}^{\top}}}\right)^{-1}\textbf{\ensuremath{\boldsymbol{X}\boldsymbol{X}^{\top}}}\right]\ast\boldsymbol{1}_{M\times N}\left(\tilde{\boldsymbol{y}}-\tilde{\boldsymbol{U}}\hat{\boldsymbol{c}}\right)\\
 & =\left\{ \left[\frac{1}{\lambda M}\boldsymbol{I}_{N\times N}-\frac{1}{\left(\lambda M\right)^{2}}\boldsymbol{X}^{\top}\left(\boldsymbol{I}_{N\times N}+\frac{1}{\lambda M}\textbf{\ensuremath{\boldsymbol{X}\boldsymbol{X}^{\top}}}\right)^{-1}\textbf{\ensuremath{\boldsymbol{X}}}\right]\boldsymbol{X}^{\top}\ast\boldsymbol{1}_{M\times N}\right\} \left(\tilde{\boldsymbol{y}}-\tilde{\boldsymbol{U}}\hat{\boldsymbol{c}}\right),
\end{align*}
with the second line following from the associativity of the Khatri-Rhao
product: $\boldsymbol{A}\ast\boldsymbol{1}_{M\times N}+\boldsymbol{B}\ast\boldsymbol{1}_{M\times N}=\left(\boldsymbol{A}+\boldsymbol{B}\right)\ast\boldsymbol{1}_{M\times N},$
and because $\left(\boldsymbol{A}\ast\boldsymbol{1}_{M\times N}\right)\boldsymbol{B}=\left(\boldsymbol{AB}\right)\ast\boldsymbol{1}_{M\times N}.$
In the last line, we pulled out $\boldsymbol{X}^{\top}$ at the right-hand
side of the $\left[\right]$ brackets. We then recognize the Woodbury
identity 
\[
\frac{1}{\lambda M}\boldsymbol{I}_{N\times N}-\frac{1}{\left(\lambda M\right)^{2}}\boldsymbol{X}^{\top}\left(\boldsymbol{I}_{N\times N}+\frac{1}{\lambda M}\textbf{\ensuremath{\boldsymbol{X}\boldsymbol{X}^{\top}}}\right)^{-1}\textbf{\ensuremath{\boldsymbol{X}}}=\left(\boldsymbol{X}^{\top}\boldsymbol{X}+\lambda M\boldsymbol{I}_{p\times p}\right)^{-1},
\]
 which finally yields
\[
\underset{\alpha\rightarrow\infty}{\lim}\hat{\boldsymbol{\beta}}=\left[\left(\boldsymbol{X}^{\top}\boldsymbol{X}+\lambda M\boldsymbol{I}_{p\times p}\right)^{-1}\boldsymbol{X}^{\top}\left(\tilde{\boldsymbol{y}}-\tilde{\boldsymbol{U}}\hat{\boldsymbol{c}}\right)\right]\ast\boldsymbol{1}_{M\times N},
\]
with $\hat{\boldsymbol{c}}$ given by (\ref{eq:AlphaLimitClinical}).
We again recognize the standard ridge regression estimator with unpenalized
$\tilde{\boldsymbol{U}}$ and with penalty $\lambda M\boldsymbol{I}_{p\times p}.$
Each entry $j$ of this estimator is repeated $M$ times because of
the Khatri-Rhao product of the standard ridge estimator with $\boldsymbol{1}_{M\times N}.$ 

\section{Regularization paths}
\label{sec:Regularization-paths}

To evaluate the effect of fusion penalty $\alpha$ on estimates of
the leaf-node-specific omics effects $\hat{\boldsymbol{\beta}},$
we show regularization plots for several fixed values of $\lambda$
($\lambda=\left\{ 0,1,10,100,500,5000\right\} $). We do so for a
simulated data set in which some omics covariates $\boldsymbol{x}_{i}$
interact with clinical covariates $\boldsymbol{z}_{i}.$ The effect
of the clinical covariates on the response is defined by a tree structure. 

\begin{figure}[H]
\centering{}\includegraphics[width=\textwidth]{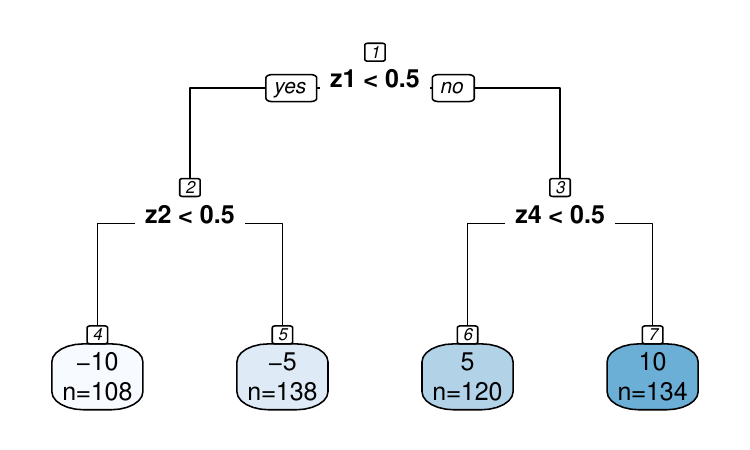}\caption{\protect\label{fig:TreeEffMod}Fit of the tree}
\end{figure}

We consider sample size $N=500$ and number of omics covariates $p=10.$
We simulate omics covariates $\boldsymbol{x_{i}}\sim\mathcal{N}\left(\boldsymbol{0}_{p},\boldsymbol{I}_{p\times p}\right),$
clinical covariates $z_{il}\sim\textrm{Unif}\left(0,1\right),$ and
define response $y_{i}=f\left(\boldsymbol{z}_{i},\boldsymbol{x}_{i}\right)+\epsilon_{i},$
with $\epsilon_{i}\sim\mathcal{N}\left(0,1\right),$ for $i=1,\ldots,N$
and clinical covariate index $l=1,\ldots,5.$ The relationship $f\left(\cdot\right)$
between clinical and omics covariates and response $y_{i}$ is given
by
\begin{align}
f\left(\boldsymbol{x},\boldsymbol{z},\boldsymbol{\beta}\right) & =I\left(z_{1}\leq\frac{1}{2}\right)I\left(z_{2}\leq\frac{1}{2}\right)\left(-10+6\boldsymbol{x}_{1,2}^{\top}\boldsymbol{\beta}_{1,2}\right)+I\left(z_{1}\leq\frac{1}{2}\right)I\left(z_{2}>\frac{1}{2}\right)\left(-5+3\boldsymbol{x}_{1,2}^{\top}\boldsymbol{\beta}_{1,2}\right)\nonumber \\
 & +I\left(z_{1}>\frac{1}{2}\right)I\left(z_{4}\leq\frac{1}{2}\right)\left(5+\frac{1}{2}\boldsymbol{x}_{1,2}^{\top}\boldsymbol{\beta}_{1,2}\right)+I\left(z_{1}>\frac{1}{2}\right)I\left(z_{4}>\frac{1}{2}\right)\left(10+\frac{1}{5}\boldsymbol{x}_{1,2}^{\top}\boldsymbol{\beta}_{1,2}\right)\nonumber \\
 & +\boldsymbol{x}_{3:10}^{\top}\boldsymbol{\beta}_{3:10.},\label{eq:RegPathInteraction}
\end{align}
with $\beta_{j}\sim\mathcal{N}\left(0,5/p\right).$ Thus, omics covariates
$j=\left\{ 1,2\right\} $ interact with the clinical covariates and
the other $8$ omics covariates do not. The estimated tree structure,
using R package \href{https://cran.r-project.org/web/packages/rpart/index.html}{\texttt{rpart}},
is shown in Figure \ref{fig:TreeEffMod}, and equals the true tree
structure specified in (\ref{eq:RegPathInteraction}).
We then estimate omics effects $\hat{\boldsymbol{\beta}}$as a function
of penalty $\alpha$ for the $\lambda$ grid. We show estimates for
$\beta_{2}=0.62,$ which interacts with $\boldsymbol{z}_{i},$ and
$\beta_{6}=-0.92,$ which does not interact with $z_{i}.$ In addition,
we show the estimated constant $c_{6}$ in node $6$ (Figure \ref{fig:TreeEffMod}),
whose true value equals $c_{6}=5.$
\begin{figure}[H]
\centering{}\includegraphics[width=\textwidth]{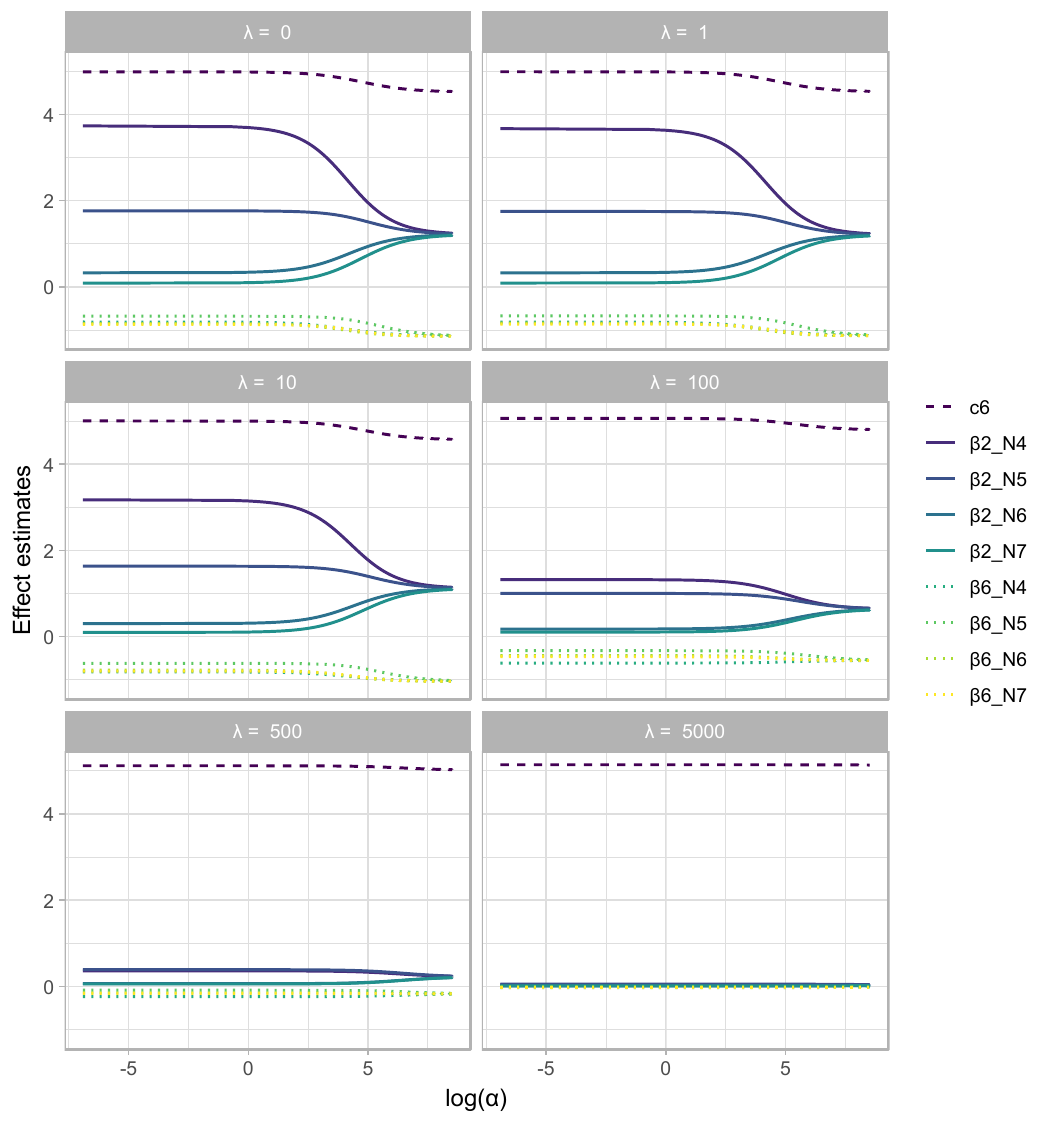}\caption{\protect\label{fig:TreeFit_RegPaths_EffMod}Regularization plots as
a function of fusion penalty $\alpha$ for several values of $\lambda$.
For illustration purposes, we only depict the node-specific estimates
of $\beta_{2},$ $\beta_{6}$ and the clinical intercept estimated
in node $6,$ i.e. $c_{6}.$}
\end{figure}

Results are depicted in Figure (\ref{fig:TreeFit_RegPaths_EffMod}).
For small $\alpha$, and $\lambda<500,$ the node-specific estimates
of $\beta_{2}$ vary substantially, which is expected because there
is a strong interaction effect between this omics covariate and the
clinical covariates. Estimates of $\beta_{6}$ remain relatively stable
across nodes, which is also expected as for this omics covariate no
interactions are present. For large $\alpha,$ the node-specific effects
of $\beta_{2}$ are shrunken towards a shared value. Figure (\ref{fig:TreeFit_RegPaths_EffMod})
also shows that larger $\lambda$ values shrink the node-specific
omics effect estimates towards $0,$ as expected. The clinical intercept
$c_{6},$ which is left unpenalized slightly decreases for large $\alpha.$
Because $c$ and $\beta$ are estimated jointly, penalization of $\beta$
by the fusion penalty introduces bias in estimation of $c$ (see limit
(\ref{eq:AlphaLimitClinical})) This bias becomes smaller for larger
$\lambda$ and diminishes for $\lim_{\lambda\rightarrow\infty},$
i.e. limit (\ref{eq:LamLimit}). 

\section{Global Test summary}
\label{sec:Global-Test}

We shortly summarize the global test methodology (\citealp{Globtest})
applied to FusedTree. In node $m,$ we have data $\mathcal{D}_{m}=\left\{ y_{k}^{(m)},\boldsymbol{x}_{k}^{(m)}\right\} _{k=1}^{n_{m}}$
and we model the response by:
\[
E\left(y_{k}^{(m)}\mid\boldsymbol{x}_{i}^{(m)}\right)=c_{m}+\sum_{j=1}^{p}x_{kj}^{(m)}\beta_{j}^{(m)}.
\]
We then test: 
\[
H_{0}:\quad\beta_{1}^{(m)}=\beta_{2}^{(m)}=\dots=\beta_{p}^{(m)}=0,
\]
which is infeasible using a standard F-test for $p>n_{m}.$ To make
progress, it is assumed that elements of $\textbf{\ensuremath{\boldsymbol{\beta}_{(m)}}}$
come from a common distribution with zero mean and variance $\tau^{2}.$
The method then tests
\[
H_{0}:\quad\tau^{2}=0,
\]
using the score test statistic. Because this statistic is asymptotically
normal under $H_{0}$, p-values may be computed from this asymptotic
distribution. Alternatively, for small sample sizes, the empirical
distribution for the test statistic may be determined using permutations.
The global test method also applies to binary $y_{i}\in\left\{ 0,1\right\} $
and survival response. For full details, see (\citealp{GlobalTest}).

\section{Simulations results}
\label{sec:Simulations-results}

Here, we show the full descriptions and results of the simulations
summarized in Section $4$ of the main text. 

We conduct three simulation experiments with different functional
relationships $f_{1},$ $f_{2},$ $f_{3}$ between the response $y$
and clinical $\boldsymbol{z}$ and omics covariates $\boldsymbol{x}$
to showcase FusedTree:

\begin{enumerate}
\item Interaction (Section \ref{subsec:Effect-modification}). We specify
$f_{1}$ inspired by model (1) of the main text. Thus $f_{1}$ is a tree,
defined by clinical covariates, with different linear omics models
in the leaf nodes for $25\%$ of the omics covariates. The remaining
$75\%$ of the omics covariates has a constant effect size. Thus,
the clinical covariates interact with $25\%$ of the omics covariates. 
\item Full Fusion (Section \ref{subsec:Full-Fusion}). In this experiment,
we specify $f_{2}$ by two separate parts, a nonlinear clinical part
and a linear omics part. In this experiment, the clinical covariates
do not act as effect modifiers and FusedTree would benefit from a
large fusion penalty $\alpha.$
\item Linear (Section \ref{subsec:Linear}). In this experiment, we specify
$f_{3}$ by a separate linear clinical and a linear omics part. Again,
FusedTree would benefit from a large fusion penalty $\alpha.$
\end{enumerate}

The set-up for the three experiments is as follows. We simulate response
$y_{i}=f\left(\boldsymbol{z}_{i},\boldsymbol{x}_{i}\right)+\epsilon_{i},$
with $\epsilon_{i}\sim\mathcal{N}\left(0,1\right)$ for $i=1,\ldots,N,$
and with different $f\left(\cdot\right)$ for each experiment. We
consider two simulation settings: $N=100$ and $N=300.$ For each
experiment and for each setting, we simulate clinical covariates $\boldsymbol{z}_{il}\sim\textrm{Unif}\left(0,1\right),$
for $l=1,\ldots,q$ and $q=5,$ and omics covariates $\boldsymbol{x}_{i}\sim\mathcal{N}\left(\boldsymbol{0}_{p},\boldsymbol{\Sigma}_{p\times p}\right),$
with $p=500,$ and correlation matrix $\boldsymbol{\Sigma}_{p\times p}$
set to the estimate of a real omics data set (\citealp{best2015rna}) of
which we randomly select $p=500$ covariates. For correlation matrix
estimation, we employ work by \citet{Schafer_2005} implemented in
the R package \href{https://cran.r-project.org/web/packages/corpcor/index.html}{\texttt{corpcor}}.
Finally, we simulate elements $j=1,\ldots,p$ of the omics effect
regression parameter vector by $\beta_{1},\ldots,\beta_{p}\sim\textrm{Laplace}(0,\theta),$
with scale parameter $\theta.$ The Laplace distribution is the prior
density for Bayesian lasso regression and ensures many close-to-zero
effect sizes. We tune $\theta$ to control the signal in the omics
covariates. Specifics of this parameter are found in the subsections.

In each experiment and for each setting, we simulate $500$ data sets.
To each data set, we fit FusedTree and several competitors: ridge
regression and lasso regression with unpenalized $\boldsymbol{z}_{i}$
implemented in the R package \href{https://cran.r-project.org/web/packages/porridge/index.html}{\texttt{porridge}} (\citealp{PorrRidge})
and \href{https://cran.r-project.org/web/packages/glmnet/index.html}{\texttt{glmnet}}\cite{GlmnetLinLog},
respectively, random forest (RF) implemented in the R package \href{https://cran.r-project.org/web/packages/randomForestSRC/index.html}{\texttt{randomforestSRC}} (\citealp{Ishwaran2008}),
and gradient boosting (\citealp{GBFriedman}) (GB) implemented in the R
package \href{https://cran.r-project.org/web/packages/gbm/index.html}{\texttt{gbm}} (\citealp{GBM}).
To assess the benefit of tuning fusion penalty $\alpha,$ we also
fit FusedTree with $\alpha=0$ (ZeroFus), and Fully FusedTree (FulFus).
Fully FusedTree jointly estimates a separate clinical part, defined
by the estimated tree, and a separate omics part that does not vary
with respect to the clinical covariates. For experiment $1$, we also
include an oracle tree model. This model knows the tree structure
in advance and only estimates the regression parameters in the leaf
nodes and tunes $\lambda$ and $\alpha$. For all FusedTree-based
models, we also include all continuous clinical covariates $\boldsymbol{z}_{i}$
linearly in the regression model, as explained in Section 2.6 of the main text.

We compare prediction models using the prediction mean square error
(PMSE): $N^{-1}\sum_{i=1}^{N}\left(y_{i}-\hat{y}_{i}\right)^{2},$
with $\hat{y}_{i}$ the prediction of the given model for observation
$i.$ The PMSE is estimated on an independent test data set of size
$N_{\textrm{test}}=5,000.$ We summarize the PMSEs over the $500$
simulated data sets using boxplots. Finally, we tune the hyperparameter
of all considered prediction models by $5$-fold cross validation.
For FusedTree, we first prune the tree and then tune $\lambda$ and
$\alpha,$ for ridge and lasso regression, we tune the standard penalties,
and for gradient boosting, we tune the learning rate and the number
of trees. We do not tune random forest because it is relatively robust
to different hyperparameter settings. 

\subsection{Interaction between clinical and omics covariates}
\label{subsec:Effect-modification}

We specify the relationship $f_{1}$ between response and clinical
and omics covariates by 
\begin{align*}
f_{1}\left(\boldsymbol{x},\boldsymbol{z},\boldsymbol{\beta}\right) & =I\left(z_{1}\leq\frac{1}{2}\right)I\left(z_{2}\leq\frac{1}{2}\right)\left(-10+8\boldsymbol{x}_{1:125}^{\top}\boldsymbol{\beta}_{1:125}\right)\\
 & +I\left(z_{1}\leq\frac{1}{2}\right)I\left(z_{2}>\frac{1}{2}\right)\left(-5+2\boldsymbol{x}_{1:125}^{\top}\boldsymbol{\beta}_{1:125}\right)\\
 & +I\left(z_{1}>\frac{1}{2}\right)I\left(z_{4}\leq\frac{1}{2}\right)\left(5+\frac{1}{2}\boldsymbol{x}_{1:125}^{\top}\boldsymbol{\beta}_{1:125}\right)\\
 & +I\left(z_{1}>\frac{1}{2}\right)I\left(z_{4}>\frac{1}{2}\right)\left(10+\frac{1}{8}\boldsymbol{x}_{1:125}^{\top}\boldsymbol{\beta}_{1:125}\right)\\
 & +\boldsymbol{x}_{126:500}^{\top}\boldsymbol{\beta}_{126:500.}+3z_{3}.
\end{align*}
The scale parameter of the Laplace distribution equals $\theta=10/p.$
Clinical covariates $\boldsymbol{z}$ contain one noise covariate:
$z_{5},$ and $4$ predictive covariates: tree covariates $z_{1},$
$z_{2},$ and $z_{4}$ and linear covariate $z_{3}.$ The predictive
clinical covariates interact with the first $25\%$ of omics covariates,
while the last $75\%$ of omics covariates have a constant effect
size. 
\begin{figure}[H]
\centering{}\includegraphics[bb=0bp 15bp 504bp 216bp,clip,width=\textwidth]{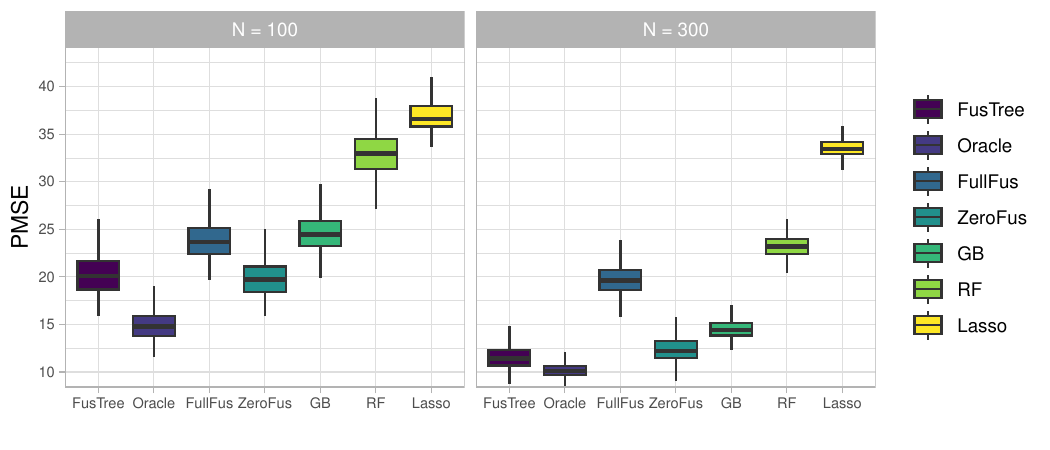}\caption{\protect\label{fig:BoxplotsEffMod}Boxplots of prediction mean square
errors for several learners across $500$ simulated data sets for
$N=100$ and $N=300$ for the interaction simulation experiment. For
illustration purposes, we excluded ridge regression because it performed
much worse compared to the other models.}
\end{figure}

FusedTree clearly outperforms FulFus and competitors GB, RF, and lasso
for both sample size settings (Figure \ref{fig:BoxplotsEffMod} and
Table \ref{tab:AvePMSE_Interaction}). We excluded results for ridge
regression because it performed much worse than the other models.

\begin{figure}[H]
\centering{}\includegraphics[width=\textwidth]{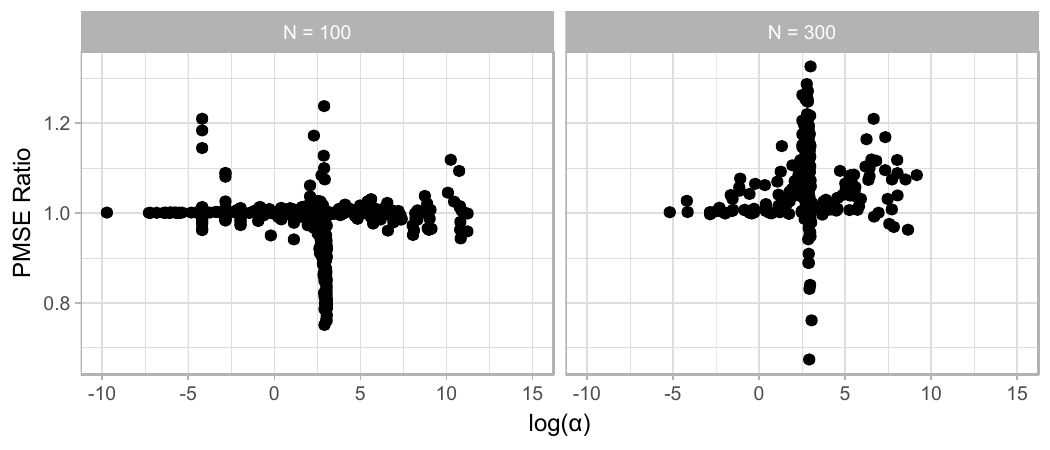}\caption{\protect\label{fig:Ratio_vs_Alpha_EffMod}Scatter plot of $\textrm{PMSE}_{\textrm{ZeroFus}}/\textrm{PMSE}_{\textrm{FusedTree}}$
as a function of fusion penalty $\alpha$ (log scale) across $500$
simulated data sets for $N=100$ and $N=300$ for the effect modification
simulation experiment (Section 4.1)}
\end{figure}

The oracle model performs better than FusedTree indicating that the
tree structure is not always estimated reliably. This difference becomes
smaller for a larger sample size because tree structure estimation
improves. For $N=100,$ FusedTree with $\alpha=0$ (ZeroFus) has a
slightly lower average PMSE than FusedTree, while FusedTree has a
lower PMSE than ZeroFus for $N=300.$ Supplementary Figure \ref{fig:Ratio_vs_Alpha_EffMod}
reveals that for $N=100,$ fusion penalty parameter $\alpha$ is in
some cases rather large, which explains why ZeroFus performs slightly
better. For $N=300,$ $\alpha$ is tuned more reliably by FusedTree.
Consequently, FusedTree has a lower average PMSE than ZeroFus. 
\begin{table}[H]
\centering{}\caption{\protect\label{tab:AvePMSE_Interaction}Average PMSE for several learners
for the interaction simulation experiment}
\begin{tabular}{ccc}
 & \textbf{$\boldsymbol{N=100}$} & \textbf{$\boldsymbol{N=300}$}\tabularnewline
\hline 
FusedTree & $20.5$ & $11.6$\tabularnewline
Oracle & $14.9$ & $10.2$\tabularnewline
FulFus & $24.1$ & $19.8$\tabularnewline
ZeroFus & $20.0$ & $12.5$\tabularnewline
GB & $24.7$ & $14.5$\tabularnewline
RF & $33.0$ & $23.2$\tabularnewline
Ridge & $50.6$ & $47.2$\tabularnewline
Lasso & $37.5$ & $33.5$\tabularnewline
\end{tabular}
\end{table}

\subsection{Full Fusion}
\label{subsec:Full-Fusion}
For the full fusion experiment, we specify $f_{2}$ by 
\begin{equation}    f_{2}\left(\boldsymbol{x},\boldsymbol{z},\boldsymbol{\beta}\right)=15\sin\left(\pi z_{1}z_{2}\right)+10\left(z_{3}-\frac{1}{2}\right)^{2}+2\exp\left(z_{4}\right)+2z_{5}+\boldsymbol{x}^{\top}\boldsymbol{\beta}.
\end{equation}
We set the scale parameter of the Laplace distribution to $\theta=75/p,$
which ensures that the clinical covariate part explains slightly more
variance in the response compared to the omics covariate part. 
\begin{figure}[H]
\centering{}\includegraphics[bb=0bp 15bp 504bp 216bp,clip,width=\textwidth]{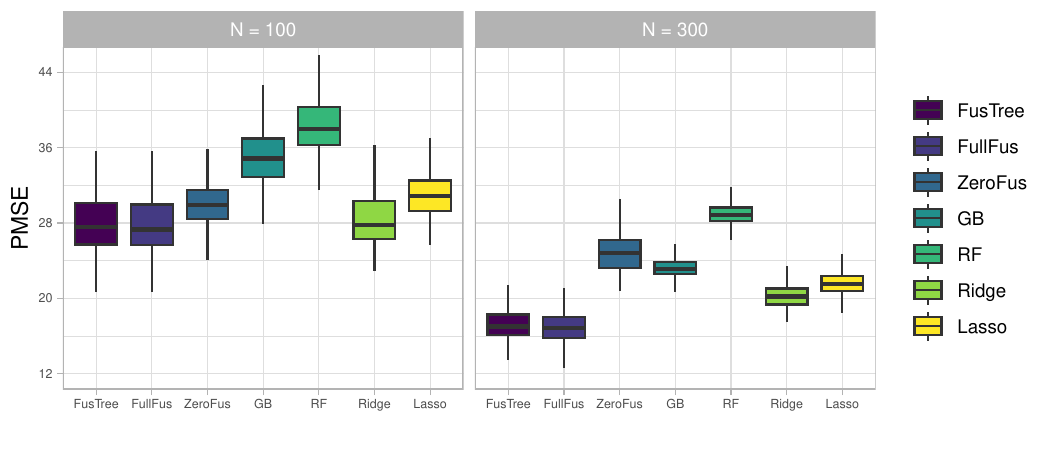}\caption{\protect\label{fig:BoxplotsFullFus}Boxplots of prediction mean square
errors for several learners across $500$ simulated data sets for
$N=100$ and $N=300$ for the full fusion simulation experiment}
\end{figure}

\begin{figure}[H]
\centering{}\includegraphics[width=\textwidth]{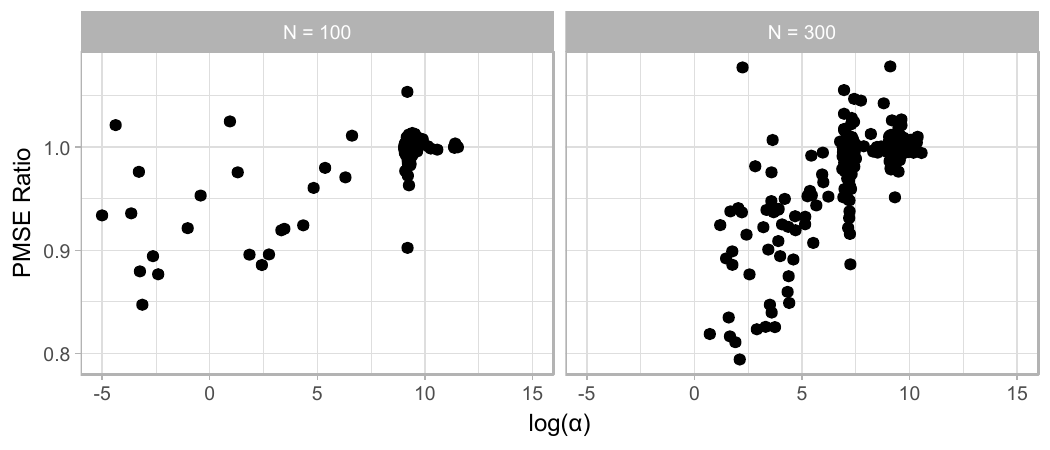}\caption{\protect\label{fig:Ratio_vs_Alpha_FullFusion}Scatter plot of $\textrm{PMSE}_{\textrm{Full\,Fus}}/\textrm{PMSE}_{\textrm{FusedTree}}$
as a function of fusion penalty $\alpha$ (log scale) across $500$
simulated data sets for $N=100$ and $N=300$ for the full fusion
simulation experiment (Section 4.2)}
\end{figure}

Figure \ref{fig:BoxplotsFullFus} and Table \ref{tab:AvePMSE_FullFusion-1}
reveal that FusedTree and FulFus perform similarly for both sample
sizes. Figure \ref{fig:Ratio_vs_Alpha_FullFusion} plots the PMSE
ratio of FulFus and FusedTree across simulated data sets as a function
of the tuned fusion penalty $\alpha.$ This plot shows that $\alpha$
is typically set to a large value in which case the ratio is close
to $1.$ For the few cases that the tuned $\alpha$ is small, FulFus
outperforms FusedTree. FusedTree has a lower PMSE compared to ZeroFus,
especially for $N=300.$ This finding suggests a clear benefit of
borrowing information across nodes compared to independently estimating
the omics effects. 

FusedTree has a lower PMSE compared to competitors ridge and lasso
regression, random forest, and gradient boosting, although the difference
with ridge and lasso regression is small for $N=100.$
\begin{table}[H]
\centering{}\caption{\protect\label{tab:AvePMSE_FullFusion-1}Average PMSE for several
learners for the full fusion simulation experiment}
\begin{tabular}{ccc}
 & \textbf{$\boldsymbol{N=100}$} & $\boldsymbol{N=300}$\tabularnewline
\hline 
FusedTree & $27.9$ & $17.4$\tabularnewline
FulFus & $27.8$ & $17.2$\tabularnewline
ZeroFus & $30.0$ & $24.8$\tabularnewline
GB & $35.0$ & $23.3$\tabularnewline
RF & $38.2$ & $29.0$\tabularnewline
Ridge & $28.2$ & $20.5$\tabularnewline
Lasso & $31.1$ & $21.6$\tabularnewline
\end{tabular}
\end{table}

\subsection{Linear}
\label{subsec:Linear}

For the linear experiment, we specify $f_{3}$ by 
\[
f_{3}\left(\boldsymbol{x},\boldsymbol{z},\boldsymbol{c},\boldsymbol{\beta}\right)=\boldsymbol{z}^{\top}\boldsymbol{c}+\boldsymbol{x}^{\top}\boldsymbol{\beta},
\]
with elements $c_{l}\sim\textrm{Laplace}\left(\frac{75}{p}\right)$
of clinical regression parameter $\boldsymbol{c}\in\mathbb{R}^{5},$
and elements $\beta_{j}\sim\textrm{Laplace}\left(\frac{35}{p}\right)$
of omics regression parameter vector $\boldsymbol{\beta}\in\mathbb{R}^{500}.$
The linear clinical part explains slightly more variation in $y$
compared to the linear omics part. 
\begin{figure}[H]
\centering{}\includegraphics[width=\textwidth]{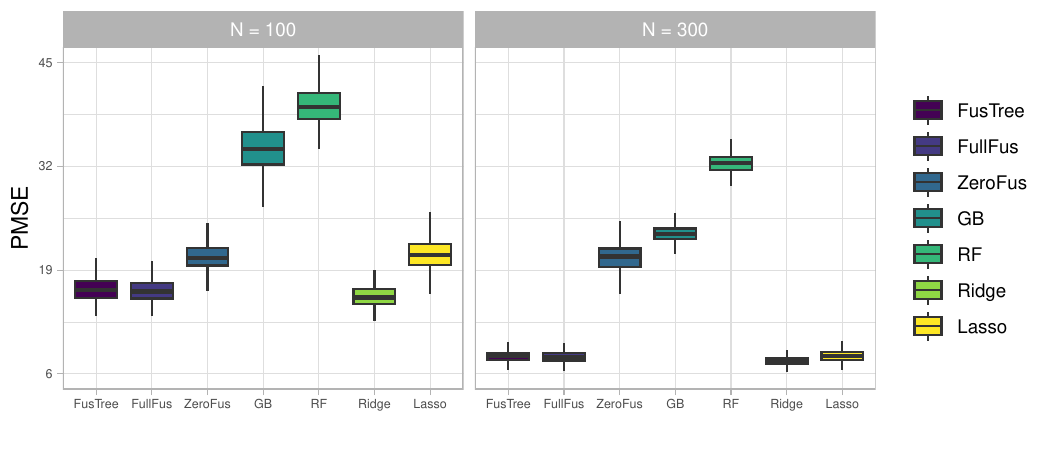}\caption{\protect\label{fig:BoxplotsLinear-1}Boxplots of prediction mean square
errors for several learners across $500$ simulated data sets for
$N=100$ and $N=300$ for the linear simulation experiment}
\end{figure}

FusedTree clearly outperforms the nonlinear competitors GB and RF
(Figure \ref{fig:BoxplotsLinear-1} and Table \ref{tab:AvePMSE_Linear-1}).
For both sample size settings, FusedTree has a slightly larger PMSE
compared to ridge regression. This difference becomes smaller for
$N=300$ compared to $N=100.$ FusedTree performs slightly better
than lasso regression for $N=100$, while for $N=300$ performance
is similar.

Compared to FulFus, FusedTree performs nearly identical. Therefore,
the benefit of fully fusing the omics effects in advance compared
to estimating the fusion strength is negligible for this simulation
experiment. Again, FusedTree has a lower PMSE than ZeroFus because
of the benefit of borrowing information across nodes.
\begin{table}[H]
\centering{}\caption{\label{tab:AvePMSE_Linear-1}Average PMSE for several learners
for the linear simulation experiment}
\begin{tabular}{ccc}
 & $\boldsymbol{N=100}$ & $\boldsymbol{N=300}$\tabularnewline
\hline 
FusedTree & $16.7$ & $8.23$\tabularnewline
FulFus & $16.5$ & $8.17$\tabularnewline
ZeroFus & $20.8$ & $20.3$\tabularnewline
GB & $34.6$ & $23.6$\tabularnewline
RF & $39.6$ & $32.5$\tabularnewline
Ridge & $15.8$ & $7.62$\tabularnewline
Lasso & $21.0$ & $8.29$\tabularnewline
\end{tabular}
\end{table}
\begin{figure}[H]
\centering{}\includegraphics[width=\textwidth]{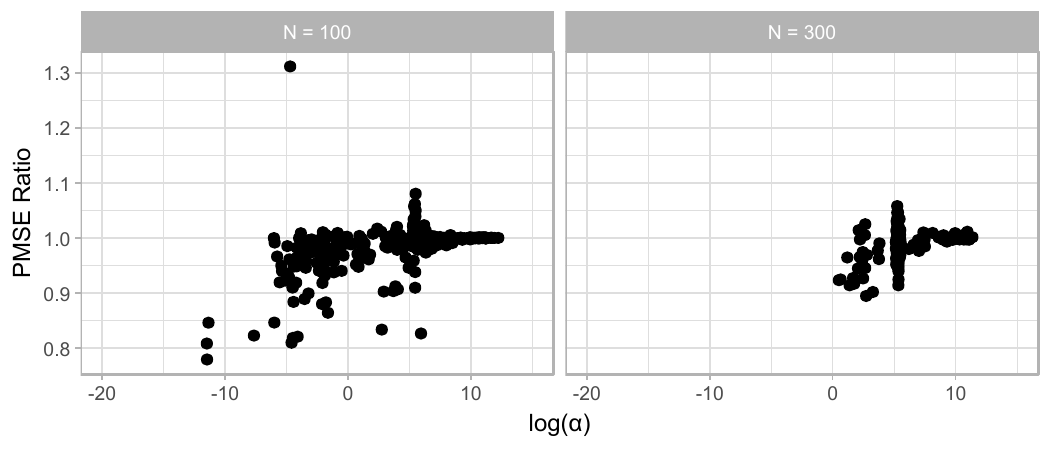}\caption{\protect\label{fig:Ratio_vs_Alpha_Linear}Scatter plot of $\textrm{PMSE}_{\textrm{Full\,Fus}}/\textrm{PMSE}_{\textrm{FusedTree}}$
as a function of fusion penalty $\alpha$ (log scale) across $500$
simulated data sets for $N=100$ and $N=300$ for the linear simulation
experiment (Section 4.3)}
\end{figure}

\section{Survival curves for CRC application}
\label{sec:Survival curves for CRC application}

\begin{figure}[H]
\centering{}\includegraphics{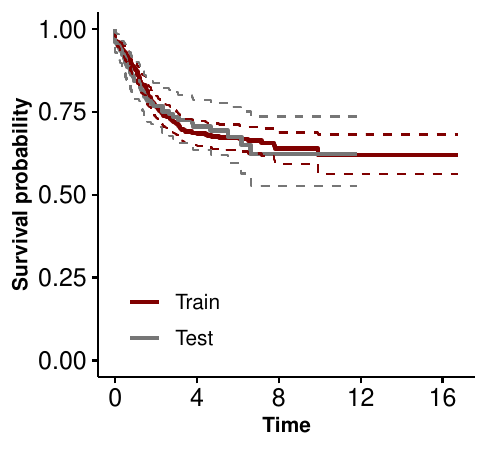}\caption{\label{fig:SurvCurve_CRC} Kaplan-Meier estimate of the overall survival probability of the training and test response
as a function of time (in years). The plot is produced using the R
package \href{https://cran.r-project.org/web/packages/survminer/index.html}{\texttt{survminer}}}
\end{figure}